\RequirePackage{fix-cm}
\RequirePackage{amsmath, bbm}
\documentclass[epj,nopacs]{svjour}
\pdfoutput=1
\smartqed  
\usepackage{tabularx} 
\usepackage{graphicx}
\usepackage{axodraw2}
\usepackage{amscd}
\usepackage{slashed}
\usepackage{multirow}
\usepackage{float}
\makeatletter
\def\cl@chapter{\@elt {theorem}}
\makeatother
\usepackage{amsmath}
\usepackage{bm}
\usepackage{cleveref}
\usepackage[caption=false]{subfig}

\newcommand{\bsmm}{{$b \rightarrow s \mu^+ \mu^-$}}
\journalname{Eur. Phys. J. C}
\begin{document}
\title{Hide and Seek with the Third Family Hypercharge Model's $Z^\prime$ at the Large
  Hadron Collider}

\titlerunning{Hide and Seek With the TFHM's $Z^\prime$ at the LHC}     

\author{B.C. Allanach \and Hannah Banks}
\institute{DAMTP, University of Cambridge, Wilberforce Road, Cambridge, CB3 0WA, United Kingdom \label{addr1}
}



\date{Received: date / Accepted: date}

\abstract{The Third Family Hypercharge Model predicts a $Z^\prime$ gauge boson with
  flavour dependent couplings which has been used to explain anomalies in
  meson decay processes which involve the
  $b\rightarrow s\mu^+\mu^-$ transition. The model predicts that a TeV-scale
  $Z^\prime$ 
  should decay to particle-antiparticle pairs of muons, taus, top quarks and bottom quarks with appreciable 
  branching ratios. We reinterpret various ATLAS and CMS search limits for
  $Z^\prime$ production followed by such decays at the LHC over a
  parameter space of the model that results from a successful combined fit to the
  $b\rightarrow s l^+ l^-$ data and precision electroweak observables. Current
  exclusions in parameter space and expected sensitivities
  of the various different channels in the high-luminosity run are compared.
  We find that the high-luminosity
  run of the LHC will significantly increase the 
  sensitivity to the model beyond the existing empirical limits, which we
  find to be surprisingly weak.
}

\maketitle
\section{Introduction \label{sec:intro}}
The Third Family Hypercharge Model  (TFHM)~\cite{Allanach:2018lvl} explains 
some data involving bottom 
quark to strange quark anti-muon muon transitions \bsmm{}
(and the charge conjugated version). Several
different measurements (dubbed `\bsmm{} anomalies')
from $B$ meson decays involving such transitions have been
shown to collectively disagree with Standard Model (SM) predictions.
In particular, measurements of the ratios of branching ratios ($BR$s)~\cite{Aaij:2021vac}
$$R_K^{(\ast)}=BR(B \rightarrow
K^{(\ast)} \mu^+ \mu^-)/BR(B \rightarrow K^{(\ast)},
e^+e^-)$$
angular distributions in $B\rightarrow K^{(\ast)} \mu^+ \mu^-$ decays~\cite{Aaij:2013qta,Aaij:2015oid,Aaboud:2018krd,Sirunyan:2017dhj,Khachatryan:2015isa,Bobeth:2017vxj},
$BR(B_s\rightarrow \mu^+\mu^-)$~\cite{Aaboud:2018mst,Chatrchyan:2013bka,CMS:2014xfa,Aaij:2017vad,LHCbtalk} and
$BR(B_s\rightarrow \phi \mu^+ \mu^-)$~\cite{Aaij:2015esa,CDF:2012qwd} show varying levels of
discrepancy with their SM predictions. The lepton flavour universality (probed here
between the electron and muon flavours of lepton) of the electroweak
interactions predicted by the SM appears to be broken.
One estimate puts the recent combined global significance of this tension
at 4.3 standard
deviations~\cite{Lancierini:2021sdf} after the look elsewhere effect and
theoretical uncertainties in the predictions have been taken into account. 

In order to ameliorate this tension, several extensions of the SM have been
proposed which invoke a spontaneously broken gauged
$U(1)$ family symmetry~\cite{Altmannshofer:2014cfa,Crivellin:2015mga,Crivellin:2015lwa,Crivellin:2015era,Altmannshofer:2015mqa,Sierra:2015fma,Celis:2015ara,Greljo:2015mma,Falkowski:2015zwa,Chiang:2016qov,Boucenna:2016wpr,Boucenna:2016qad,Ko:2017lzd,Alonso:2017bff,Tang:2017gkz,Bhatia:2017tgo,Fuyuto:2017sys,Bian:2017xzg,Alonso:2017uky,Bonilla:2017lsq,King:2018fcg,Duan:2018akc,Kang:2019vng,Calibbi:2019lvs,Altmannshofer:2019xda,Capdevila:2020rrl,Davighi:2020qqa,Allanach:2020kss,Bednyakov:2021fof,Davighi:2021oel,Greljo:2021npi,Wang:2021uqz}. The broken family symmetry yields a
massive $Z^\prime$ spin 1 boson with family-dependent couplings. Provided that
such a
boson has a coupling to $\bar b s + \bar s b$  quarks and to $\mu^+ \mu^-$, it can explain
the \bsmm{} anomalies via the process
depicted in Fig.~\ref{fig:bsmumu}. 
\begin{figure}
\begin{center}
  \begin{axopicture}(80,55)(-5,-5)
    \Line[arrow](25,25)(50,50)
    \Line[arrow](0,25)(25,25)
    \Line[arrow](75,50)(50,25)
    \Line[arrow](50,25)(75,0)
    \Photon(25,25)(50,25){3}{3}
    \Text(37.5,17)[c]{$Z^{\prime}$}
    \Text(-5,25)[c]{$b$}
    \Text(45,50)[c]{$s$}
    \Text(82,50)[c]{$\mu^+$}
    \Text(82,0)[c]{$\mu^-$}    
  \end{axopicture}
\caption{\label{fig:bsmumu} A $Z^\prime$-boson mediated
  contribution to \bsmm{} transitions.}
\end{center}
\end{figure}
However, the measurement of $B_s-\overline{B_s}$ mixing is more in line with its SM prediction
and so the contribution to $B_s-\overline{B_s}$ mixing from the $Z^\prime$-mediated process in
Fig.~\ref{fig:bsbs} receives an upper bound. 
\begin{figure}
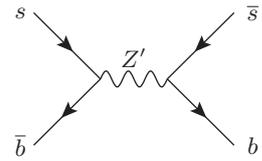

\begin{center}
  \begin{axopicture}(80,55)(-5,-5)
    \Line[arrow](0,50)(25,25)
    \Line[arrow](25,25)(0,0)
    \Line[arrow](75,50)(50,25)
    \Line[arrow](50,25)(75,0)
    \Photon(25,25)(50,25){3}{3}
    \Text(37.5,33)[c]{$Z^\prime$}
    \Text(-5,0)[c]{$\overline{b}$}
    \Text(-5,50)[c]{$s$}
    \Text(82,50)[c]{$\overline{s}$}
    \Text(82,0)[c]{$b$}    
  \end{axopicture}
\caption{\label{fig:bsbs} A $Z^\prime$-boson mediated
  contribution to $B_s - \overline{B_s}$ mixing.}
\end{center}
\end{figure}
The combination of fitting the \bsmm{} anomalies as well as
$B_s-\overline{B}_s$ mixing measurements then implies that the $Z^\prime$ coupling to
$\bar b s$ quarks, although non-zero,
is much smaller than its coupling to $\mu^+\mu^-$. This can happen
automatically in models where the $Z^\prime$ couples only to third family
quarks in the weak eigenbasis, but where a small mixing produces a coupling to
$\bar bs + \bar s b$ in the mass eigenbasis. This is the \emph{modus operandi} of 
the TFHM~\cite{Allanach:2018lvl}.

\subsection{The model}

We shall now introduce the TFHM model, which is the main subject of
study of
the present paper. The initial gauge group is
$SU(3)\times SU(2)_L \times U(1)_Y \times U(1)_{Y_3}$ and the field content of
the model, shown in Table~\ref{tab:content}, is minimally extended from the
SM\@. Indeed, gauge singlet fermions can be added (with no other changes to the
model) in order to provide explicitly for 
neutrino masses and mixings. Since neutrino masses and mixings do not play a
r{o}le in our analysis, we neglect such gauge singlet fermions. 
\renewcommand{\arraystretch}{1.2} 
\begin{table}
\begin{center}
\begin{tabular}{|c|ccccc|}\hline
field & spin & $SU(3)$ & $SU(2)_L$ & $U(1)_Y$ & $U(1)_{Y_3}$ \\
\hline
$G^{\prime\mu}$ & 1 & 8 & 1 & 0 & 0 \\
$W^{\prime\mu}$ & 1 & 1 & 3 & 0 & 0 \\
$B^{\prime\mu}$ & 1 & 1 & 1 & 0 & 0 \\
$X^{\prime\mu}$ & 1 & 1 & 1 & 0 & 0 \\
\hline
$Q_{L_i}^\prime$ & $\frac{1}{2}$ & 3 & 2 & $\frac{1}{6}$ & 0 \\
$u_{R_i}^\prime$ & $\frac{1}{2}$ & 3 & 1 & $\frac{2}{3}$ & 0 \\
$d_{R_i}^\prime$ & $\frac{1}{2}$ & 3 & 1 & $-\frac{1}{3}$ & 0 \\
$L_{L_1}^\prime$ & $\frac{1}{2}$ & 1 & 2 & $-\frac{1}{2}$ & 0 \\
$e_{R_1}^\prime$ & $\frac{1}{2}$ & 1 & 1 & $-1$ & 0 \\
$L_{L_2}^\prime$ & $\frac{1}{2}$ & 1 & 2 & $-\frac{1}{2}$ & $-\frac{1}{2}$ \\
$e_{R_2}^\prime$ & $\frac{1}{2}$ & 1 & 1 & $-1$ &  $0$ \\
$Q_{L_3}^\prime$ & $\frac{1}{2}$ & 3 & 2 & $\frac{1}{6}$ & $\frac{1}{6}$ \\
$u_{R_3}^\prime$ & $\frac{1}{2}$ & 3 & 1 & $\frac{2}{3}$ & $\frac{2}{3}$ \\
$d_{R_3}^\prime$ & $\frac{1}{2}$ & 3 & 1 & $-\frac{1}{3}$ & $-\frac{1}{3}$ \\
$L_{L_3}^\prime$ & $\frac{1}{2}$ & 1 & 2 & $-\frac{1}{2}$ & $0$ \\
$e_{R_3}^\prime$ & $\frac{1}{2}$ & 1 & 1 & $-1$ &  $-1$ \\
\hline
$H$ & 0 & 1 & 1 & $\frac{1}{2}$ & $\frac{1}{2}$ \\
$\theta$ & 0 & 1 & 1 & 0 & 1 \\
\hline \end{tabular}
\end{center}
\caption{Field content of the TFHM and representations under the gauge
group. $i \in \{1,2\}$ is a family index for the 
first two families. The left-handed doubles further decompose as
$Q_{L_\alpha}^\prime:=(u_{L_\alpha}^\prime,\ d_{L_\alpha}^\prime)^T$ and
$L_{L_\alpha}^\prime:=(\nu_{L_\alpha}^\prime,\ \nu_{L_\alpha}^\prime)^T$,
respectively, where 
$\alpha \in \{1,2,3\}$.
Here, all fields are listed in the (primed) gauge eigenbasis and
the hypercharge gauge boson $B^{\prime \mu}$ and $Y_3$ gauge boson ($X^{\mu\prime}$) fields are
defined in a basis in which 
there is no kinetic mixing between them.\label{tab:content}}
\end{table}
\renewcommand{\arraystretch}{1}
$\theta$ is the complex scalar flavon field, whose vacuum
expectation value $\langle \theta \rangle$ is assumed to be around the TeV
scale, breaking $U(1)_{Y_3}$. The 
$X^{\prime \mu}$ gauge boson acquires a TeV-scale mass term through `eating' the goldstone boson
\begin{equation}
M_{X} = g_X \langle \theta \rangle,
\end{equation}
where $g_X$ is the gauge coupling of the $U(1)_{Y_3}$ group.
Since the Higgs doublet $H$ has a non-zero $U(1)_{Y_3}$ quantum number, its
vacuum expectation value $v$ induces mass squared-term
mixing between the hypercharge boson $B^{\mu\prime}$, $X^{\mu\prime}$ and 
the electrically neutral component of the $W^{\mu\prime}$.
This `$Z^0-Z^\prime$ mixing' alters the SM prediction of the electroweak
observables~\cite{Allanach:2018lvl}. 
The three gauge boson mass eigenstates that result
are the (massless) photon, the $Z^0$ 
boson of mass $M_Z$ and the $Z^\prime$ boson of mass $M_{Z^\prime}$. The two non-zero gauge boson masses are
\begin{align}
M_{Z^\prime} =& M_X\left[ 1 + {\mathcal
O}\left( \frac{M_Z^2}{M_{Z^\prime}^2}\right)\right], \nonumber \\
M_Z =& v \frac{\sqrt{g^2+{g^\prime}^2}}{2} \left[1+ {\mathcal
O}\left( \frac{M_Z^2}{M_{Z^\prime}^2}\right)\right],
\end{align}
at tree-level,
where $g^\prime$ and $g$ are the gauge couplings of $U(1)_Y$ and $SU(2)_L$, respectively.

The mixing between the weak eigenbasis (primed) fields and the mass eigenbasis
(unprimed) fermionic fields is given by
\begin{equation}
{\bm F}^\prime = V_F {\bm F},
\end{equation}
where
\begin{equation}
F \in \{u_L,\ d_L,\ \nu_L,\ e_L,\ u_R,\ d_R,\ e_R\},
\end{equation}
and ${\bm F}:=( F_1,\ F_2,\ F_3 )^T$ is thought of as a column 3-vector in
family space and $V_F$ are unitary 3 by 3 matrices. After such mixing, the CKM
quark-mixing matrix and the PMNS lepton-mixing matrices are
\begin{equation}
V = V_{u_L}^\dag V_{d_L}, \qquad
U = V_{\nu_L}^\dag V_{e_L}, \label{mixings}
\end{equation}
respectively.
The original example case
of the TFHM's assumptions, which we shall adhere to here, correspond 
to: $V_{u_R}=V_{d_R}=V_{e_R}=V_{e_L}=I_3$, the 3 by 3 identity matrix.
The example case allows $b_L-s_L$ mixing, i.e.\
\begin{equation}
V_{d_L}= \begin{pmatrix}
1 & 0 & 0 \\
0 & \cos \theta_{23} & \sin \theta_{23} \\
0 & -\sin \theta_{23} & \cos \theta_{23} \\
\end{pmatrix},
\end{equation}
in order to facilitate the necessary beyond-the-SM flavour changing
for \bsmm{} transitions. $\theta_{23}$ is considered to be a parameter of
the model and is varied.
$V_{u_L}$ and $V_{\nu_L}$ are then fixed by (\ref{mixings}) by inputting the
central empirically-inferred values for $U$ and
$V$~\cite{ParticleDataGroup:2020ssz}. 

\subsection{The fit}

\begin{figure}
\begin{center}
\includegraphics[width=\columnwidth]{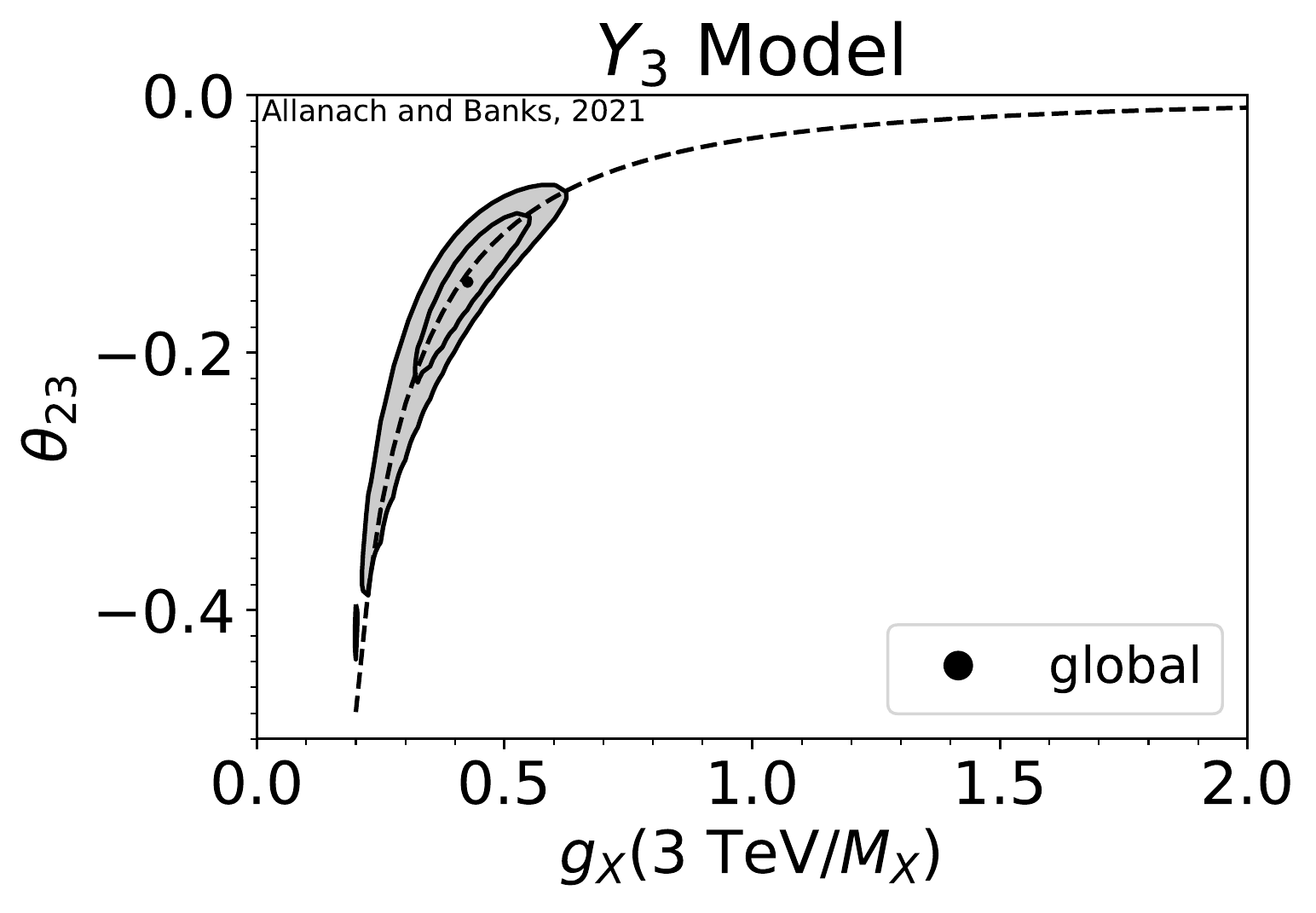}
\end{center}
\caption{Fit of electroweak and \bsmm{} data to the TFHM\@. This plot was made from the data
presented in the ancillary information of the {\tt arXiv} version of
Ref.~\cite{Allanach:2021kzj}. The dot displays the best-fit point and the
inner (outer) contours in the shaded region display the 68\% (95\%) confidence
level (CL) contours. The dashed line shows our characterisation of the fit. \label{fig:y3fit}}
\end{figure}

In Ref.~\cite{Allanach:2021kzj}, it was shown that the TFHM
with two fitted parameters $x:=g_X(\text{3~TeV}/M_X)$ and $\theta_{23}$ can fit
219 combined data from the \bsmm{} and electroweak
sectors better than the SM by $\sqrt{\Delta \chi^2}=6.5\sigma$. This fit utilises
the SM effective field theory in order to predict the observables used. This is
only a good approximation for the electroweak observables if $M_{Z^\prime} \gg
M_Z$. We display the result of the fit in
Fig.~\ref{fig:y3fit}.

The relevant TFHM parameters for us are $M_X$, $g_X$ and $\theta_{23}$. It
will suit us to reduce these by one parameter by characterising the main
variation of the fit. We do this by restricting 
\begin{equation}
\theta_{23} = \frac{-0.09}{x^2 + 0.35x}, \label{t23}
\end{equation}
shown by the dashed line in Fig.~\ref{fig:y3fit}.
The domain of good fit (i.e.\ within the 95$\%$ CL region) corresponds to
\begin{equation}
x \in [0.2,\ 0.6]. \label{good_fit}
\end{equation}
We thus reduce the number of 
effective parameters to two: $x$ and $M_X$. In practice, we note that up to terms of
order $M_{Z^2 / M_X^2}$, $x=g_{Z^\prime}/M_{Z^\prime}$ and $M_X=M_{Z^\prime}$. By
taking our two input parameters to be
$g_{Z^\prime}\times \text{3~TeV}/M_{Z^\prime}$ and $M_{Z^\prime}$, we are thus able to
perform two dimensional parameter scans over the parameter space which, to a good
approximation, characterises the
region of good fit over the domain in (\ref{good_fit}). 

Armed with a parameterisation of the well-fit parameter space, it is of
interest to ask whether one may directly observe the $Z^\prime$ in scattering
experiments, since this would constitute a `smoking gun' signal of the model
(and others of its ilk). 

\subsection{Previous work on TFHM $Z^\prime$ searches}
\begin{figure}
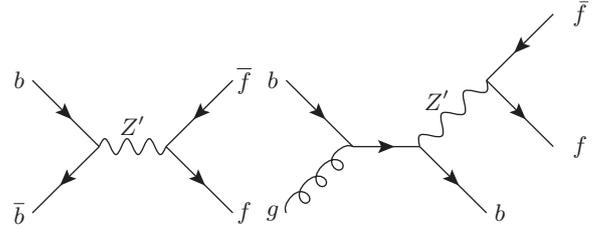

\begin{center}
  \begin{axopicture}(200,75)(-25,-5)
    \Line[arrow](-20,50)(5,25)
    \Line[arrow](5,25)(-20,0)
    \Line[arrow](55,50)(30,25)
    \Line[arrow](30,25)(55,0)
    \Photon(5,25)(30,25){3}{3}
    \Text(17.5,33)[c]{$Z^\prime$}
    \Text(-25,0)[c]{$\overline{b}$}
    \Text(-25,50)[c]{$b$}
    \Text(62,50)[r]{$\overline{f}$}
    \Text(62,0)[r]{$f$}
    
\Gluon(75,0)(100,25){3}{3}
\Line[arrow](75,50)(100,25)
\Line[arrow](100,25)(125,25)
\Line[arrow](125,25)(150,0)
\Photon(125,25)(150,50){3}{3}
\Line[arrow](175,75)(150,50)
\Line[arrow](150,50)(175,25)
\Text(70,50)[c]{$b$}
\Text(70,0)[c]{$g$}
\Text(155,0)[c]{$b$}
\Text(185,75)[c]{$\bar f$}
\Text(185,25)[c]{$f$}
\Text(131.5,42.5)[c]{$Z^\prime$}
\end{axopicture}
\end{center}
\caption{\label{fig:ffbar} LHC production of a $Z^\prime$, followed by its
decay into a fermion anti-fermion pair $f \bar f$. In the right-hand panel, we
show the associated production with a $b$ quark.}
\end{figure}

Previous work reinterpreting LHC searches for the $Z^\prime$ predicted by
the TFHM is on the $Z^\prime \rightarrow \mu^+\mu^-$ channel. In some sense
this is natural because first-two family leptons are cleaner objects
experimentally and because the model predicts a sizeable branching ratio in
this channel, as  Table~\ref{tab:BRs} shows.
\begin{table}
\begin{center}
\begin{tabular}{|cc|cc|cc|}
\hline 
Mode & $BR$ & Mode & $BR$ & Mode & $BR$ \\ \hline
$t\bar t$ & 0.42 & $b\bar b$ & 0.12 & $\nu \bar \nu'$ & 0.08 \\
$\mu^+ \mu^-$ & 0.08 & $\tau^+ \tau^-$ & 0.30 & & \\
\hline
\end{tabular}
\end{center}
\caption{\label{tab:BRs} Branching ratios of various $Z^\prime$ decays  predicted by the TFHM in
the $M_{Z^\prime} \gg 2m_t$ limit, from Ref.~\protect\cite{Allanach:2018lvl}.}
\end{table}
Ref.~\cite{Allanach:2019mfl} reinterpreted a 139 fb$^{-1}$ 13 TeV $pp$
collision ATLAS search~\cite{ATLAS:2019erb}
in the di-muon channel to find exclusion bounds upon the TFHM\@. The
determination was at tree-level and took into account processes such as the
one depicted in the left-hand panel of Fig.~\ref{fig:ffbar}. The
dominant hard production 
process is $b \bar b \rightarrow Z^\prime$ and the bounds coming from
non-observation of a significant bump in the di-muon mass spectrum are
consequently far
weaker than those in which the $Z^\prime$ couples significantly to the first two families of
quark, since they are doubly suppressed by tiny $b$ quark parton
distributions. In Ref.~\cite{Allanach:2019mfl}, the associated production
process 
depicted in the right-hand panel of Fig.~\ref{fig:ffbar} was not explicitly
simulated, but instead was taken 
into account in the approximation that the emitted $b$ quark is always soft, by
using 5-flavour parton distribution functions. Computations were performed at parton level only with no simulation of parton showering, initial state radiation or hadronisation. 

\subsection{The present paper}
The main purpose of the present paper is to exploit the other decay $Z^\prime$
channels, i.e.\ $t \bar t$, $b \bar b$, $\tau^+ \tau^-$ as well as searches
explicitly looking for associated production of di-muons with the additional
$b$ quark. We shall compare
the various constraints or sensitivities with those of the di-muon
search. In contrast to Ref.~\cite{Allanach:2019mfl}, we shall simulate parton
showering, hadronisation and initial state radiation and we
will also  explicitly take into account associated production diagrams like the
one on the right-hand panel of Fig.~\ref{fig:ffbar} using parton matching, as appropriate. This will have the additional side benefit of upgrading the
interpretation in Ref.~\cite{Allanach:2019mfl} of the
aforementioned ATLAS di-muon search~\cite{ATLAS:2019erb} with a more accurate
calculation. Finally, 
we shall overlay the exclusion regions coming from the searches
to find the overall current LHC Run II TFHM direct search constraints. 

The paper proceeds as follows: in \S\ref{sec:exp}, we detail the experimental
searches that we shall use in addition to our approximations in calculating the
exclusions. In \S\ref{sec:limits}, we present our calculated exclusion limits upon the
TFHM\@. In some cases, where there is no limit, we show the current value of the
ratio $\mu$, defined as the current 95\%CL experimental  limit divided by the expected signal
cross-section $\sigma$. This will give an indication of whether limits from
the channel 
in question can be expected in the coming years of high-luminosity
(HL)-LHC running. We 
summarise and conclude in \S\ref{sec:disc}. An estimation of theoretical
errors in the various current limits is relegated to Appendix~\ref{sec:THun}.

\section{Reinterpreting Experimental Searches \label{sec:exp}} 
In recent years, several direct searches for new spin-1 electrically neutral
resonances have been performed by both the ATLAS and CMS experiments at the
LHC for
a variety of different assumed final states. Of specific relevance to the
$Z^{\prime}$ of the TFHM are those final states consistent with the $\mu^{+} \mu^{-}$,
$\tau^{+}\tau{-}$, $b\bar{b}$ and $t\bar{t}$ decay channels, for which the
branching ratios are appreciable, as Table~\ref{tab:BRs} shows.  No search to
date has yielded a significant signal, and upper limits on
the cross-section for a putative $Z^\prime$ resonance times its branching ratio into 
each final state
have
been set, typically as a function of invariant mass of the resonance, $M_{Z^\prime}$. In what follows, we 
undertake a systematic re-interpretation of a number of recent $Z^\prime$ searches to
produce an up-to-date set of collider constraints on the TFHM\@. 

The generic procedure adopted for each experimental search is as follows:
\begin{enumerate}
\item At each point in the scanned parameter space, compute an estimation of
  the experimentally bounded observable within the TFHM\@.  For the searches
  considered, the relevant observables  are either a fiducial cross-section (
  i.e.\ accounting for experimental acceptance) or 
  the total $Z^{\prime}$ production cross-section,  each times a final state branching ratio.
\item Plot the locus of the experimental bound in $g_{Z^\prime} \times
  3\text{~TeV}/M_{Z^\prime}$ for each value of $M_{Z^\prime}$ being
  considered. $\theta_{23}$ is fixed by (\ref{t23}).
\end{enumerate}
A number of the searches that we consider publish bounds for a set of
different values of the $Z^{\prime}$ 
width-to-mass ratio, $\Gamma / M_{Z^\prime}$. Let us denote this set as $\{W_1,W_2, \ldots ,W_n\}$ where $W_j > W_i$ for $j > i$ and $W_1$ corresponds to the   $\Gamma \rightarrow 0$ limit.  We express the corresponding experimental bounds as $\{B(W_1,M_{Z^\prime}),B(W_2,M_{Z^\prime}), \ldots , B(W_n,M_{Z^{\prime}})\} $. 
For a \linebreak generic $z \equiv \Gamma / M_{Z^\prime}$ such that  $W_p < z < W_{p+1}$
we take the bound to be the linear interpolation between 
$\ln(B(W_p,M_{Z^\prime}))$ and  $\ln(B(W_{p+1},M_{Z^\prime}))$:
\begin{equation}
B(z,M_{Z^\prime}) =  B(W_{p},M_{Z^\prime})\left(\frac{B(W_{p+1},M_{Z^\prime})}{B(W_{p},M_{Z^\prime})}\right)^{\left(\frac{z - W_{p}}{W_{p+1} -W_{p}}\right)}.
\end{equation}

\begin{figure}

    \centering
    \includegraphics[width=\columnwidth]{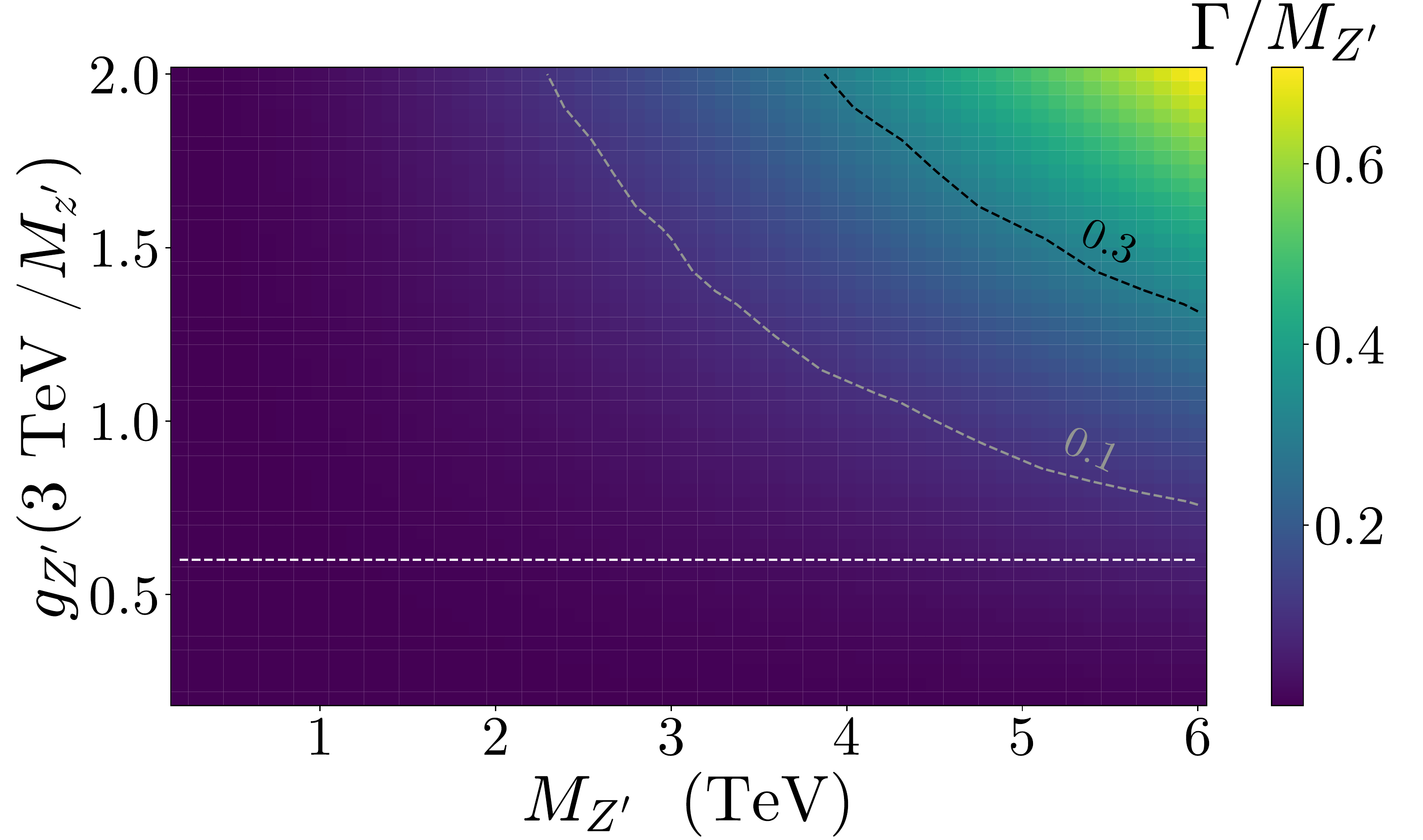}
    \caption{The width-to-mass ratio, $\Gamma /  M_{Z^\prime}$ of the TFHM
      $Z^{\prime}$ boson over the scanned parameter space. The contours
      $\Gamma /  M_{Z^\prime}$  = 0.1 and  $\Gamma /  M_{Z^\prime}$  = 0.3  are shown by the dashed grey and black lines respectively. The
      region favoured by the flavour and electro-weak data falls between the
      dashed white line and the bottom of the plot. 
}
    \label{fig:w_m} 
\end{figure}

 For $z > W_n$ we set $B(z,M_{Z^\prime})$ = $B(W_n,M_{Z^\prime})$ but caution the reader of this
approximation by delineating these regions of parameter space in our plots.
It is important to note that at $\Gamma/M_{Z^\prime} \sim {\mathcal O}(1)$, the TFHM becomes non-perturbative
and our results can no longer be considered reliable, since they are based on
perturbative calculations. {The $Z^\prime$ propagator that we use misses relative
corrections of ${\mathcal O}(\Gamma^2/M_{Z^\prime}^2)$ and so 
we might expect ${\mathcal O}(10)\%$
relative corrections to the cross-section when $\Gamma/M_{Z^\prime}\geq 0.3$.}
As shown in Fig.~\ref{fig:w_m}
however, for the vast majority of the scanned parameter space, and the
parameter region favoured by the flavour and electroweak data, $\Gamma /
M_{Z^\prime} $ is well below 0.3.
Neglecting $Z - Z^{\prime}$ mixing, the $Z^\prime$ width-to-mass ratio in the
limit of zero fermion 
masses, can be analytically approximated by \cite{Allanach:2018lvl} 
\begin{equation}
\frac{\Gamma} {M_{Z^\prime}}= \frac{5g_{Z^{\prime}}^{2}}{36 \pi}.
\end{equation}

To estimate the relevant observables, we used a Universal
Feynrules Output (UFO) file from Ref.~\cite{Allanach:2019mfl} which was obtained by
implementing the TFHM Lagrangian in {\tt
  FEYNRULES} \cite{Christensen_2009,Degrande_2012}. The {\tt UFO} file was
imported into\linebreak  {\tt MadGraph5\_aMC@NLO 
  v2.7.2 }~\cite{Alwall_2014} which was used to simulate tree-level
$Z^\prime$ production from  centre-of-mass energy $\sqrt{s}=13$~TeV proton proton
($pp$) collisions and the subsequent decay of the $Z^\prime$
for each decay channel: $pp \rightarrow Z^\prime \rightarrow X \bar{X}$  for $X \in
\{\mu, \tau,b,t\} $.  The simulated events are subsequently  fed to  {\tt Pythia8
  v2.4.5}~\cite{Sjostrand:2014zea} for the incorporation of parton showering, initial state radiation and
hadronisation effects.  We allow for the  production of a jet alongside the
$Z^{\prime}$ by explicitly including the process $pp\rightarrow Z^\prime j, Z^\prime
\rightarrow X \bar{X}$ at parton level. To avoid double-counting events with
final state jets initiated during the parton-shower with  events with final state jets originating from
the matrix element, we match up to one jet (additional to $X\bar{X}$) using the MLM procedure~\cite{2007} as implemented in  {\tt Madgraph}. We set the {\tt xqcut} parameter to be 35 GeV for the $\mu^{+} \mu^{-}$ and $b\bar{b}$ channels, and
40 GeV for the $\tau^{+} \tau^{-}$  and $t\bar{t}$ channels.
These values were  selected to produce 
smooth differential-jet-rate distributions and matched 
cross-sections which are approximately invariant to small variations of {\tt xqcut}.
All computations are carried out within a 5-flavour scheme, including the $b$ quark
in the proton and jet definitions\footnote{The masses and
  Yukawa couplings of the $b$ and light quarks are set to zero via a
  {\tt UFO} restriction file.},  using the {\tt NNPDF2.3LO}  parton distribution functions  (PDFs). Our
calculations allow for tree-level off-shell contributions  but neglect any
interference with SM backgrounds. We do not include multi-parton interactions
in the {\tt Pythia} simulations: these were found to have a negligible
influence on the relevant observables.  
   
Whilst the published bounds implicitly account for efficiencies rendering a
full detector simulation redundant, the {\tt Pythia8} output is passed to {\tt
  Delphes-3.5.0}~\cite{deFavereau:2013fsa} where jet-clustering is performed
using the {\tt fastjet-3.3.3} plugin~\cite{Cacciari:2011ma,Cacciari:2005hq}, and where relevant, $b$-tagging and muon isolation are simulated. For each of the searches, jets are clustered using the anti-$k_t$ algorithm~\cite{Cacciari:2008gp} with a distance parameter of $R = 0.4$. The resulting sets of jet and muon objects are each ordered according to transverse momentum, $p_T$, and written to a  {\tt ROOT} file. We shall refer to a member of these sets using the notation $j_i $ and $\mu_i$ respectively, where the index $i$ denotes the position in the ordered set with $i = 1$  corresponding to the ``leading object'' - that with the highest $p_T$.

The various searches that we recast differ in terms of whether or not
detector acceptance 
is included in the presented experimental bound. To
allow the same simulated events to be used across multiple analyses, we do not impose
any phase-space cuts on the events at parton level such that the matched
cross-sections for each channel computed using {\tt
  MadGraph5\_aMC@NLO } constitute our estimates for the total $Z^\prime$
production 
cross-section times branching ratio, $\sigma \times BR$.  
For searches which include acceptance in the experimental bound, we multiply each of these by the fraction, $f$, of showered events that meet the experimental selection criteria of that search. This is found by implementing the relevant phase-space cuts on the {\tt Delphes} output in the {\tt ROOT} macro.  We sum the $f \times \sigma \times BR$ from each $Z^{\prime}$ decay channel to obtain our final estimate of the experimentally bounded fiducial observable.

 In the subsequent sections, we provide additional details of the specific
 re-interpretation procedure deployed for each of the different searches that
 we recast. 
 
    \subsection{CMS di-lepton Search}
 We first re-interpret the most recent CMS search~\cite{CMS:2021ctt} for new narrow
 resonances in di-lepton (i.e.\ $e^+e^-$ and $\mu^+\mu^-$) final states using the
 full set of Run II $pp$ collision data recorded at $\sqrt{s}$ = 13 TeV.  We
 use the di-muon channel, for which the total integrated luminosity  is
 140 fb$^{-1}$. No significant signal was reported and upper limits on 
 \begin{equation}
 \label{ob}
\frac{\sigma  \times BR(Z^{\prime} \rightarrow \mu^{+} \mu^{-})}{\sigma_{Z} \times (BR(Z \rightarrow e^{+} e^{-} ) + BR(Z \rightarrow  \mu^{+} \mu^{-}))} \times 1928~\textnormal{pb},
 \end{equation} 
 where $\sigma_{Z}$ is the total production cross-section of a $Z^0$ boson, and the SM prediction of the denominator is 1928~pb.
 
 We estimate the numerator of (\ref{ob}) using our matched muon 
 channel computation, and the denominator by an analogous matched computation for the $Z^0$ boson but
 allowing for decay into either $e^+e^-$ or $\mu^+\mu^-$ pairs. In Ref.~\cite{CMS:2021ctt}, the experimental bounds were presented for $Z^{\prime}$ width-to-mass ratios of 0\%, 0.6\%, 3\%,  5\% and 10\%.   We interpolate between these as detailed above. 

\subsection{ATLAS di-muon search}
For comparison, we recast the recent ATLAS search for di-lepton resonances in
139 fb$^{-1}$  of 13 TeV $pp$ collision data~\cite{ATLAS:2019erb}, once again looking within the di-muon
channel. This bounds a fiducial cross-section times branching ratio as a
function of $Z^{\prime}$ mass for various width-to-mass ratios up to and
including 0.1. The acceptance fraction is estimated from events that contain a
pair of oppositely charged muons each with transverse momentum $p_{T} >$ 30
GeV and absolute value of pseudo-rapidity $|\eta| < 2.5$
and have a combined di-muon invariant mass $m_{\mu \mu}>$ 225 GeV. Additionally,
each muon is required to be isolated: the scalar sum of tracks with $p_{T}
> $ 1 GeV within a cone of size $\Delta R = \sqrt{(\Delta \eta)^2 + (\Delta \phi)^2 }$ to the muon, where $\Delta \eta$ is the difference in pseudo-rapidity and $\Delta \phi$ is the difference in azimuthal angle, is required to be less than 6\% of
the muon candidate's transverse momentum: $p_{T}(\mu)$. For a given muon, $\Delta
R$ is taken to be the minimum of 0.3 and the ratio 10 GeV/$p_{T}(\mu)$.  We
implement this in {\tt Delphes} by modifying the default
isolation module.   

\subsection{ATLAS di-lepton search with $b$-tagging}
\label{leptonb}
The recent ATLAS search for resonances in di-muon final
states with either exactly one or zero $b$-tagged jets is
of particular relevance to the TFHM~\cite{ATLAS:2021mla}.
The analysis was
of the full Run II dataset comprised of 139 fb$^{-1}$ of
$\sqrt{s} = 13$ TeV $pp$ collision data. 
The published bounds on the cross-section times branching ratio  include both
the acceptance and the $b$-tagging efficiency. We account for the latter via use of the default $b$-tagging module in  {\tt Delphes} to tag $b$ jets with a fixed efficiency of 77\%~\cite{ATLAS:2021mla}. To estimate the acceptance fraction, we impose identical phase space cuts to those used in the experimental selection procedure.   Jets are required to have a $p_{T} > $ 30 GeV and $|\eta| <$ 2.5 and muon candidates must have $p_{T} > $ 30 GeV and $|\eta| <$ 2. Similarly, electron candidates are required to have an energy greater than 30 GeV and  $|\eta| < $ 2.47, excluding  the region 1.37 $< |\eta| < $ 1.52. Overlap between jet and leptonic objects is removed according to the following algorithm which we effect in the {\tt ROOT} macro:

\begin{enumerate}
\item Any jet with $\Delta R <$ 0.2 with respect to an electron is
  removed. 
\item Electrons with $\Delta R <$ 0.4 with respect to a remaining jet are removed.
\item If a jet is within  $\Delta R <0.04 + 10$ GeV$/p_{T}(\mu)$ of a muon,
  it is removed if the number of constituent tracks (which we consider to be charged leptons and charged hadrons) with transverse momentum
  $p_T>0.5$ GeV is at most 2, otherwise the muon is removed. 
\end{enumerate}.

Following this procedure, events containing a pair of oppositely charged muons
are classed into two mutually exclusive categories: those with no $b$-tagged
jets ($b$-veto), and those with exactly one $b$-tagged jet ($b$-tag). For
inclusion in either class the $p_T$ of the leading muon, $p_T(\mu_1)$,
must exceed 65 GeV to
ensure selection by the trigger.  

The experimental bounds published for this search are presented as a
function of the \emph{minimum} invariant mass of the di-muon final state,
$m_{\mu \mu}$. For a given point in our parameter space,
$(M_{Z^{\prime}},g_Z^{\prime})$, we impose the experimental bound for
a value $m_{\mu \mu} = M_{Z^{\prime}} - 2 \Gamma_{Z^{\prime}}$, which provides
us with a conservative estimate of the $Z^{\prime}$ exclusion limit. Given
that the $Z^{\prime}$ width is negligible for much of parameter space, we note
that the choice of mapping between $M_Z^{\prime}$ and $m_{\mu \mu}$ used here
only has a small effect.

 \subsection{ATLAS di-jet search}
   ATLAS have released a search for new resonances in pairs of jets using 139
   fb$^{-1}$ of $pp$ collision data recorded at  $\sqrt{s}$ = 13 TeV~\cite{ATLAS:2019fgd}. Bounds are placed  on a fiducial cross-section times branching ratio for various $Z^{\prime}$ width-mass ratios of up to 0.15.  The
   di-jet system is formed from the two jets of greatest $p_T$: $j_1$ and $j_2$.
The events are categorised into an inclusive class with no
   $b$-tag requirement,  a one-$b$ tagged class   ($1b$) consisting of events
   in which at least one jet of the di-jet pair is $b$-tagged, and a doubly
   $b$-tagged class ($2b$) consisting of events in which both of the leading
   jets are $b$-tagged.  We simulate $b$-tagging using the default module in
   {\tt Delphes} setting the $b$-tag efficiency to be the product of the 77\%
   working point, with the energy dependent correction factors given in Fig.~1 of Ref. \cite{ATLAS:2019fgd}.  
      
We select the events to be included in each search category by imposing the same restrictions employed in the experimental analysis on the clustered and tagged jet objects in our {\tt
  ROOT} output files. For an event to be included in any of the categories,
the leading jet is required to have $p_{T}(j_1) >$ 420 GeV (to meet the trigger
requirements) and the secondary jet $p_T(j_2) > $ 150 GeV.  The azimuthal angle
between the jet pair $\Delta \phi (j_1j_2)$ is required to be $>$ 1. Additional,
category specific cuts are placed on $|\eta(j_1)|$ and $|\eta(j_2)|$, the invariant mass of
the di-jet system $m_{jj}$, and  half the rapidity separation of the leading
jets $y^\ast = (y(j_1) - y(j_2) )/2$, where $y$ is the rapidity.
We detail these in Table~\ref{cuts}.  

\begin{table}
\begin{center}
\begin{tabular}{|c|c|c|c|}
\hline 
Category &  Inclusive & $1b$& $2b$ \\ \hline
$ |\eta(j)|$  & - & $< 2$& $<2$ \\ 
$ |y^{\ast}|$  & $<0.6$& $< 0.8$& $<0.8$ \\ 
$ m_{jj}$  & $>1100$ GeV& $> 1133$ GeV& $> 1133$ GeV \\
$b$-tagging & -  &$\geq 1 b$-tagged jet  &$2 b$-tagged jets \\  \hline 
\end{tabular}
\end{center}
\caption{Event selection cuts imposed on the two jets of highest $p_T$ ($j_1$ and$ j_2$) for
  each analysis category in the ATLAS di-jet search~\cite{ATLAS:2019fgd}.}
\label{cuts}
\end{table}

We add the contributions from all 4 decay channels to form our overall estimate of the fiducial cross-section times branching ratio. 

\subsection{CMS di-jet search}
The CMS collaboration has also published a search for new, high mass
(i.e.\ larger than 1.8
TeV) di-jet resonances decaying to jet pairs~\cite{CMS:2019gwf}. The analysis uses $\sqrt{s} = $
13 TeV LHC $pp$ collision data corresponding to an integrated luminosity of
137 fb$^{-1}$, and bounds the fiducial cross-section times branching ratio for
a number of values of $\Gamma/M_{Z^{\prime}}$ up to 0.55.  In
order to meet the trigger requirements, either the scalar $p_T$ sum of all clustered
jets in the event with $p_T>30$~GeV and  $|\eta|< 3$ is required to exceed
1050 GeV, or at least one jet clustered with an increased distance parameter
of $R=0.8$ must have a $p_T >550$~GeV.  

Jets are required to have $p_{T}> $30 GeV and $|\eta| < $2.5.  Additionally, the energy associated with neutral hadron constituents must be less than 90\% of the                                                           total jet energy, as must the energy of constituent photons. Any jets within the fiducial tracker coverage of $|\eta| < $2.4 must have electron and muon energies of less than 90\% and 80\% of the total jet energy respectively, and must also have a non-zero energy contribution from charged hadrons~\cite{CMS:2019gwf}.  

The two jets with the largest $p_{T}$ are then taken to seed a pair of ``wide jets".  The wide jets are formed by adding  any jet within a
distance  $\Delta R <$ 1.1 to the  nearest leading jet and are designed to collect any hard-gluon radiation from the leading final-state partons. 
These wide jets are taken to form the di-jet pair such that the di-jet mass is equal to the invariant mass
of the entire wide jet system.
For an event to be included in the analysis, the wide jets must have a
pseudo-rapidity separation $|\Delta \eta | < 1.1$. In addition, the di-jet
invariant mass is required to be larger than 1.5 TeV.  We include signal contributions from all 4 possible decay channels.

\subsection{ATLAS di-tau search}
We next consider an ATLAS search \cite{Aaboud_2018} for new gauge bosons in
di-tau final states within 36 fb$^{-1}$ of $\sqrt{s} = 13$ TeV $pp$ collision
data recorded during 2015-2016. Analysis of final states  in which at least one of the taus has decayed to hadrons and neutrinos revealed no significant deviation from the SM. 95\% CL upper
limits on the production cross-section times branching ratio to di-tau final
states are presented. The experimental bound is an upper limit on $\sigma
\times BR$ so our estimation of $\sigma \times BR$ is just the matched cross-section from our {\tt Madgraph} di-tau channel calculation, with no additional selection cuts applied.  
\subsection{ATLAS di-top search}
We finally turn to an ATLAS search for new heavy particles decaying into
  ($t\bar{t}$) quark pairs~\cite{ATLAS:2020lks}. The search is performed using 139
fb$^{-1}$ of $\sqrt{s} = 13$ TeV $pp$ collisions consistent with the
production of a pair of high-transverse-momentum top quarks and their decay
into fully hadronic final states. No significant indication of an excess beyond the SM
background was found and bounds were released on the product of the
$Z^{\prime}$ production cross-section and branching fraction to $t\bar{t}$
pairs.
The experimental bound quoted in Ref.~\cite{ATLAS:2020lks} is upon $\sigma
\times BR$, so there is no need to impose any additional phase-space cuts on
our  $t\bar{t}$ channel events.  

\section{Exclusion Limits \label{sec:limits}}
For each search, we compute the parameter $\mu$, defined as the ratio of the
95\% CL experimental upper bound on the signal cross-section (either production or fiducial) times branching ratio, to the theoretical prediction of the same quantity:
\begin{equation}
\mu = \frac{\textnormal{95\% CL experimental upper bound on } \sigma \times
  BR}{\textnormal{Theoretical prediction of }\sigma \times BR}, \label{mu}
\end{equation}
at each point in parameter space. A grid interpolation over the parameter
plane is used to obtain the contour $\mu = 1$, which corresponds to the
upper-limit on the model at the 95\% confidence level. Values of $\mu>1$
indicate that 95$\%$ sensitivity has not yet been reached.

We also provide a rough and conservative estimate of  the sensitivity of the HL-LHC 
to the TFHM assuming an integrated luminosity $L$ of 3000 fb$^{-1}$ and
a centre-of-mass energy $\sqrt{s} = 13$ TeV. 
For a given centre-of-mass
energy, the number of both signal and background events are expected to scale
proportionally to the integrated luminosity of the run. Given that the number of background
events has an uncertainty of  $\pm \sqrt{L}$, one can expect the
signal-sensitivity to scale as $L/ \sqrt{L} = \sqrt{L}$. We plot a contour at
$\mu = \sqrt{3000/ L_{0}}$, where $L_{0}$ is the integrated luminosity of the present
search in question in fb$^{-1}$,  to estimate the sensitivity of using
3000 fb$^{-1}$ of $\sqrt{s} = $ 13 TeV data. Our assessment of the HL-LHC capability has two main simplifications. Firstly, the design centre-of-mass energy of future LHC runs is greater than 13 TeV. In general one expects an increase in run energy is to have a positive impact on the signal sensitivity at high
  $M_{Z^\prime}$ - our  contour
should thus be considered a conservative evaluation.  Secondly, we use the  observed upper bound in our definition of $\mu$ as opposed to the $\textit{expected}$ upper bound in the limit of infinite statistics.  
Given that the precise LHC run schedule is yet to be determined, and that  the
precise details of experimental analyses are likely to change each run, we
consider this rough estimate to be sufficient.

In the figures that follow we display $\mu$ for each of the various searches
considered. At each scan point, we plot the minimum of the
calculated value of $\mu$ and 100 (for ease of reading the plot). Where they exist, we show the obtained
exclusion limit (the contour $\mu=1$) as a solid red line, and use a
dashed red line for our HL-LHC sensitivity estimate. We show in a solid
purple line, an overall exclusion limit, obtained from a piecewise
combination of the most constraining bounds for each mass. The area
between the dashed white line and the bottom of the plot indicates the region favoured by the flavour and
electroweak data at the 95$\%$ CL\@. We shall refer to this as the `favoured region'.

\begin{figure}

    \centering
    \includegraphics[width=\columnwidth]{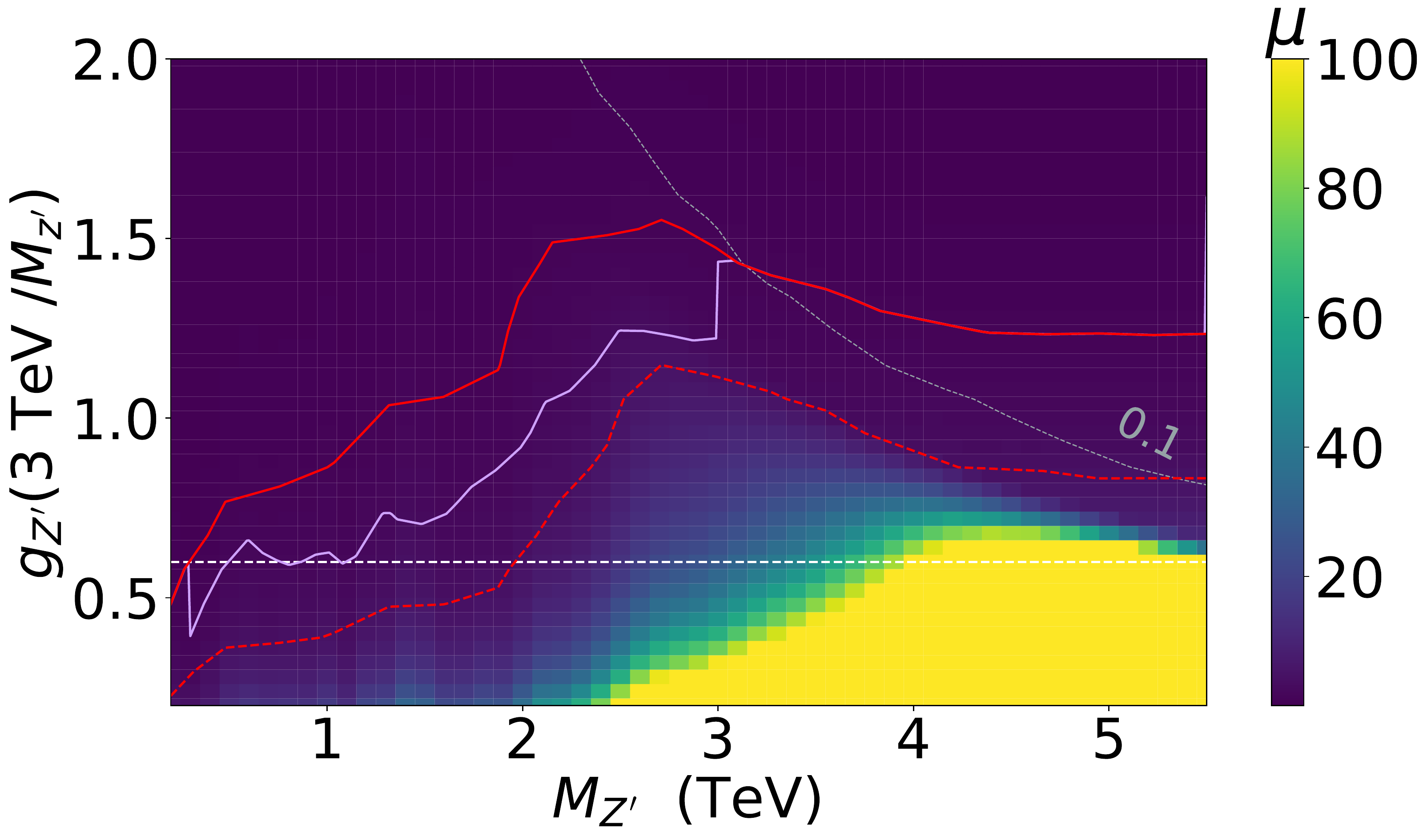}
    \caption{$\mu$ for the CMS di-muon search~\cite{CMS:2021ctt} at each point
      in the scanned parameter space. The solid red line shows the contour
      $\mu = 1$. The area above this is excluded at the 95\% CL\@. The dashed red
      line shows our estimate of the HL-LHC sensitivity. The solid purple
      line shows the current observed bound from all searches considered in this
      work. The dashed grey line shows the contour where $\Gamma /
      M_{Z^\prime}$  = 0.1, and the region between the dashed white line and
      the bottom of the plot is favoured by the flavour and electroweak data
      at the 95$\%$CL\@. Toward the extreme
      left-hand side of the plot, our predictions become more inaccurate due
      to unaccounted-for ${\mathcal O}(M_Z^2 / M_{Z^\prime}^2)$ relative corrections.
}
    \label{fig:cms_dm} 
\end{figure}

\begin{figure}

    \centering
    \includegraphics[width=\columnwidth]{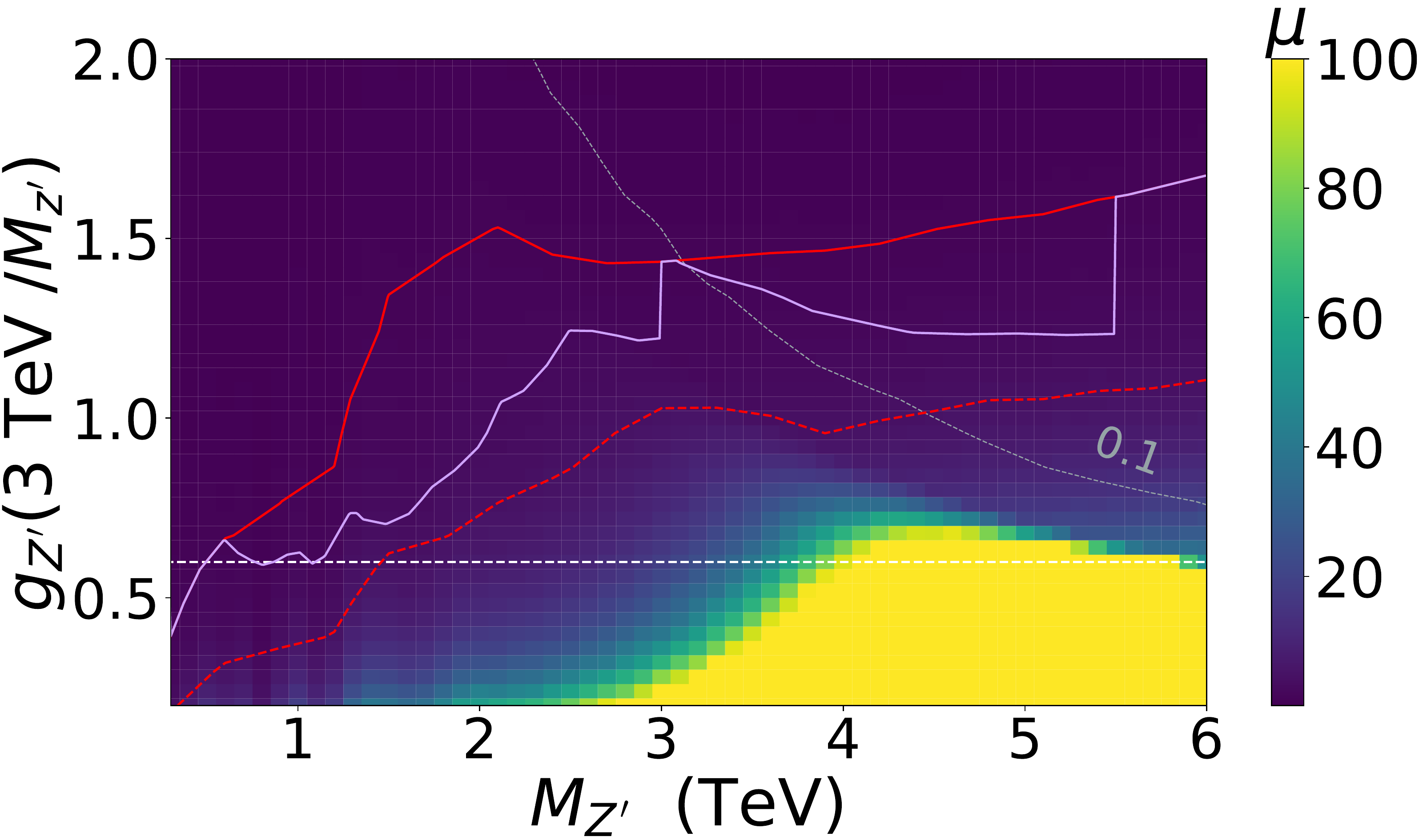}
    \caption{$\mu$ for the ATLAS di-muon search~\cite{ATLAS:2019erb} at each point in the scanned parameter space. The solid red line shows the contour $\mu = 1$. The area above this is excluded at the 95\% CL\@. The dashed red line shows our estimate of the HL-LHC sensitivity. The solid purple line shows the current observed bound from all searches considered in this work. The dashed grey line shows the contour where $\Gamma /  M_{Z^\prime}$  = 0.1, and the region between the dashed white line and the bottom of the plot is favoured by the flavour and electroweak data      at the 95$\%$CL\@. Toward the extreme
      left-hand side of the plot, our predictions become more inaccurate due
      to unaccounted-for ${\mathcal O}(M_Z^2 / M_{Z^\prime}^2)$ relative corrections.}
    \label{fig:at_dm} 
\end{figure}

We find the di-muon decay channel to be the most constraining. Figs.~\ref{fig:cms_dm} and~\ref{fig:at_dm} show the results of our reinterpretation
of the  CMS and ATLAS di-muon searches respectively. The sensitivities of the
experiments are similar, however the CMS bound is stronger for $Z^{\prime}$ masses in the range 3-5.5~TeV, where it is the most constraining search of those considered here.
  Of the experimental analyses used in the present paper, the ATLAS and CMS di-muon searches
  are the only two searches to place any constraints on the favoured region.  The constraints within the favoured region are far from severe: only a small portion of the
  $M_{Z^\prime}<0.5$ TeV parameter space is ruled out by the searches.
  We expect similar searches using HL-LHC data to be
  capable of improving the present overall exclusion limit over the entire
  mass range, with appreciable sensitivity to the favoured region for values of
  $M_{Z^\prime}$ up to about 2 TeV. 

\begin{figure}

    \centering
    \includegraphics[width=\columnwidth]{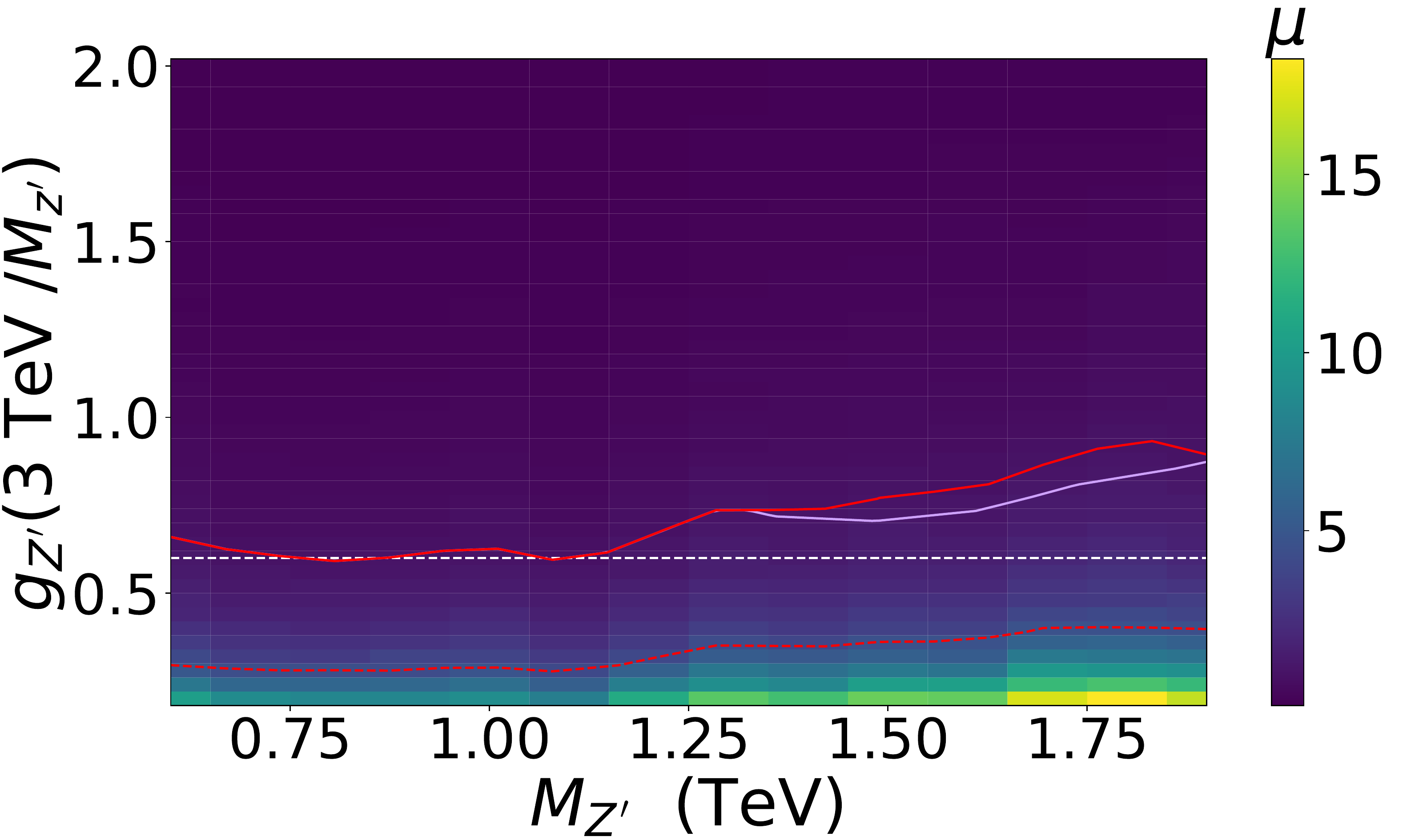}
    \caption{$\mu$ for the  $b$-tag category of the ATLAS di-muon
      search~\cite{ATLAS:2021mla} at each point in the scanned parameter space. The solid
      red line shows the contour $\mu = 1$. The area above this is excluded at the 95\% CL\@. The dashed red line shows our estimate of the HL-LHC 
      sensitivity. The solid purple line shows the current observed bound from all
      searches considered in this work. The region between the dashed white
      line and the bottom of the plot is favoured by the flavour and
      electroweak data       at the 95$\%$CL\@. Toward the extreme
      left-hand side of the plot, our predictions become more inaccurate due
      to unaccounted-for ${\mathcal O}(M_Z^2 / M_{Z^\prime}^2)$ relative corrections.}
    \label{fig:atlas_dm_btag} 
\end{figure}

\begin{figure}

    \centering
    \includegraphics[width=\columnwidth]{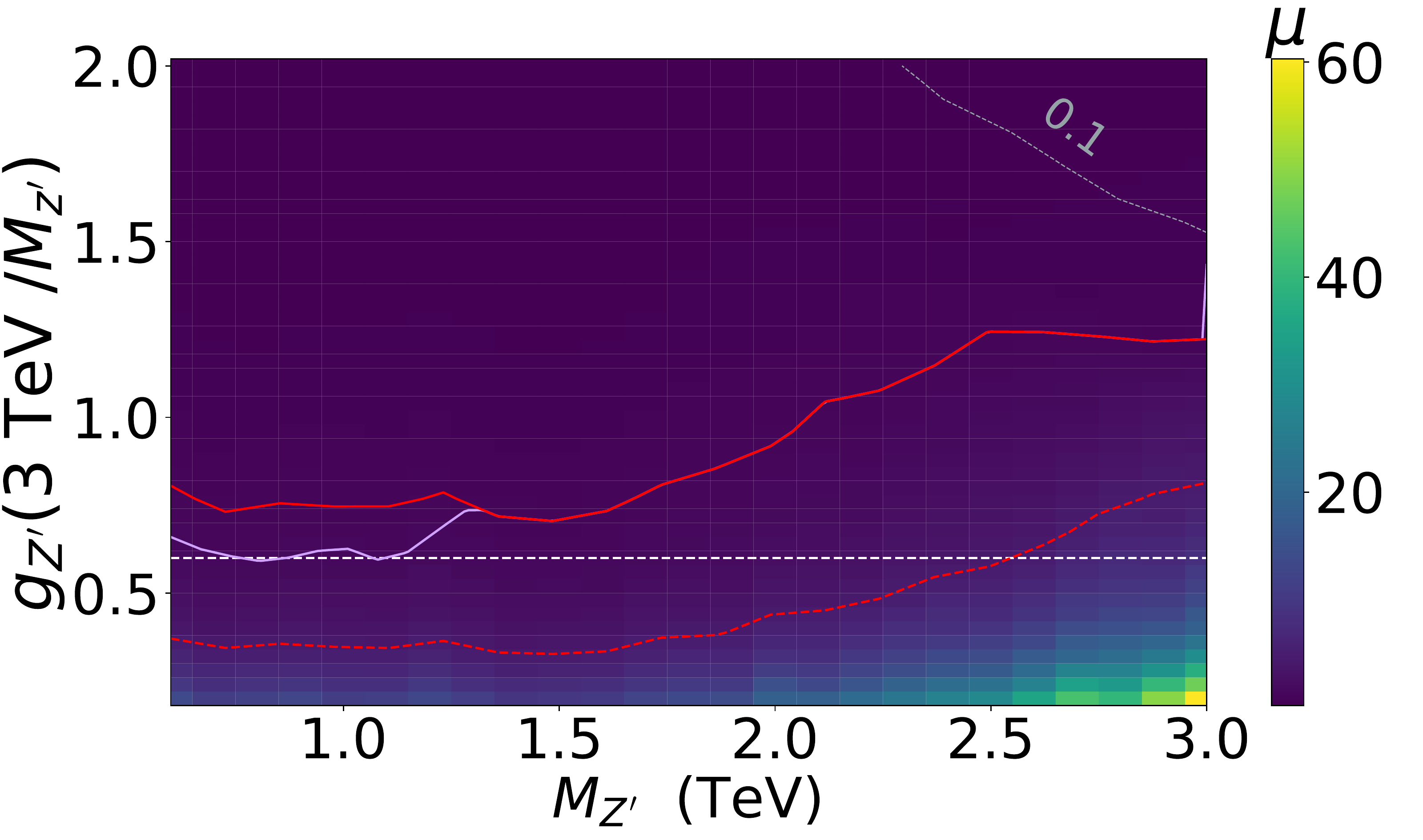}
    \caption{$\mu$ for the $b$-veto category of the ATLAS di-muon
      search~\cite{ATLAS:2021mla} at each point in the scanned parameter
      space. The solid red line shows the contour $\mu = 1$. The area above
      this is excluded at the 95\% CL\@. The dashed red line shows our estimate
      of the HL-LHC sensitivity. The solid purple line shows the current
      observed bound from all searches considered in this work. The dashed
      grey line shows the contour where $\Gamma /  M_{Z^\prime}$  = 0.1, and
      the region between the dashed white line  and the bottom of the plot is
      favoured by the flavour and electroweak data       at the 95$\%$CL\@. Toward the extreme
      left-hand side of the plot, our predictions become more inaccurate due
      to unaccounted-for ${\mathcal O}(M_Z^2 / M_{Z^\prime}^2)$ relative corrections.}
    \label{fig:atlas_dm_bveto} 
\end{figure}

For the entire 0.6 - 3 TeV mass range that it considers, the bounds from
the ATLAS di-muon analysis with explicit $b$-tagging requirements in \S\ref{leptonb} dominate
the sensitivity. The contour plots for both the $b$-tag and $b$-veto search
categories are shown in Figs.~\ref{fig:atlas_dm_btag} and~\ref{fig:atlas_dm_bveto} respectively. The $b$-tag category is most
constraining of all of the searches for 
$M_{Z^\prime}$ less than about 1.3 TeV. Above this point the $b$-veto search
becomes more 
constraining.  The HL-LHC gains sensitivity to much of the favoured region for $M_{Z^{\prime}}$ $<$ 2.5 TeV. 

The only other search providing any current constraint on the scanned
parameter space is that of the $\tau^{+}\tau^{-}$ decay channel. The present
bound is localised around $M_{Z^\prime}\sim 0.8$ TeV,  
\begin{figure}

    \centering
    \includegraphics[width=\columnwidth]{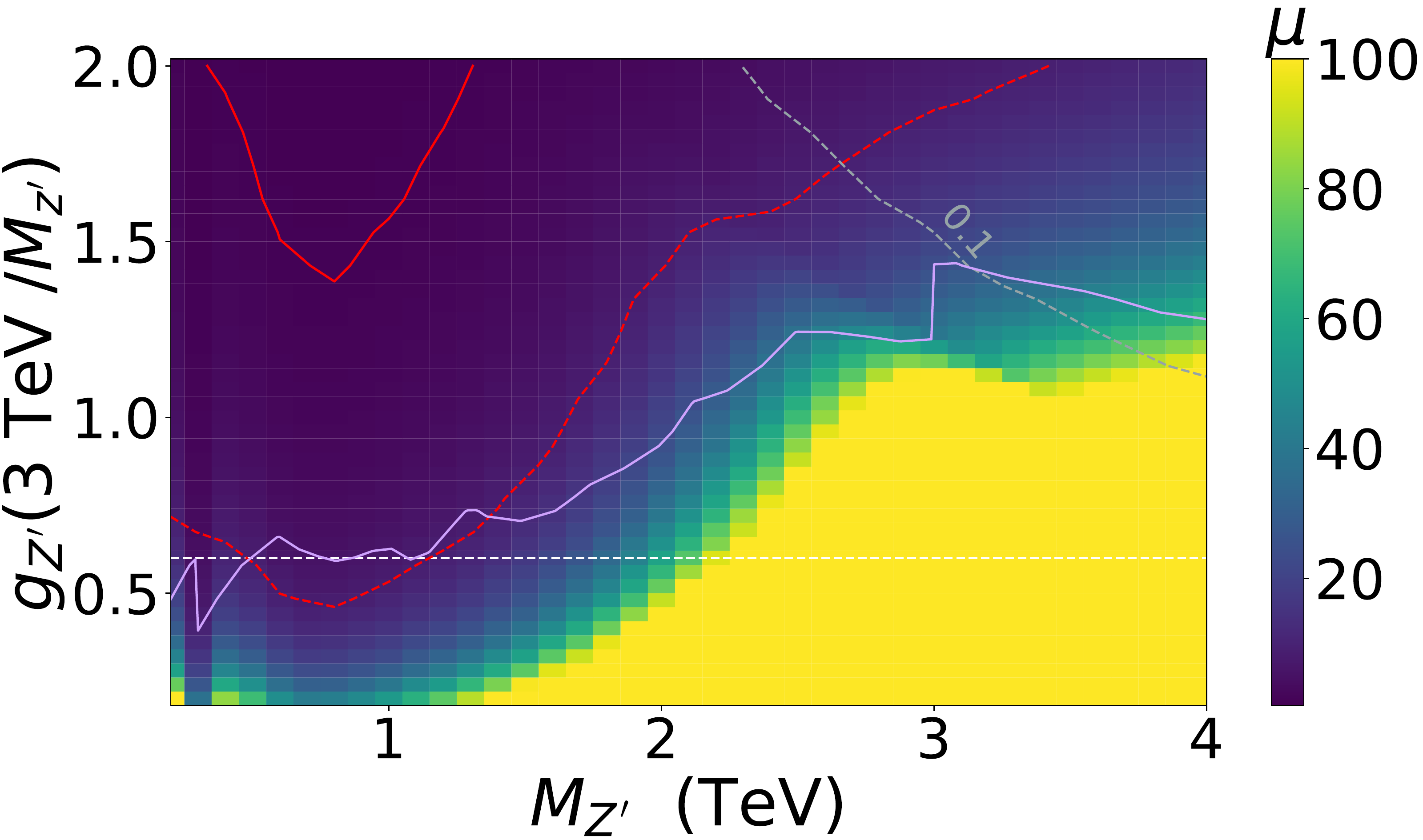}
    \caption{$\mu$ for the ATLAS di-tau~\cite{Aaboud_2018} search at each point in
      the scanned parameter space. The solid red line shows the contour $\mu =
      1$. The area above this is excluded at the 95\% CL\@.The dashed red line
      shows our estimate of the HL-LHC sensitivity. The solid purple line
      shows the current observed bound from all searches considered in this work. The dashed grey line shows the contour where $\Gamma /  M_{Z^\prime}$  = 0.1, and the region between the dashed white line and the bottom of the plot is favoured by the flavour and electroweak data      at the 95$\%$CL\@. Toward the extreme
      left-hand side of the plot, our predictions become more inaccurate due
      to unaccounted-for ${\mathcal O}(M_Z^2 / M_{Z^\prime}^2)$ relative corrections.}
    \label{fig:tau} 
\end{figure}
and is considerably weaker than the combined bound shown in Fig.~\ref{fig:tau}
(the combined bound is dominated by the di-muon channel, see
Figs.~\ref{fig:cms_dm},\ref{fig:at_dm}). Given the low integrated luminosity (36.1 fb$^{-1}$) of the experimental analysis in this channel, there is significant scope
for improved coverage at the HL-LHC, with the possibility of improving the current combined search sensitivity to $M_{Z^{\prime}}$ in the range 0.5 - 1.4 TeV.  

\begin{figure}
    \centering
    \includegraphics[width=\columnwidth]{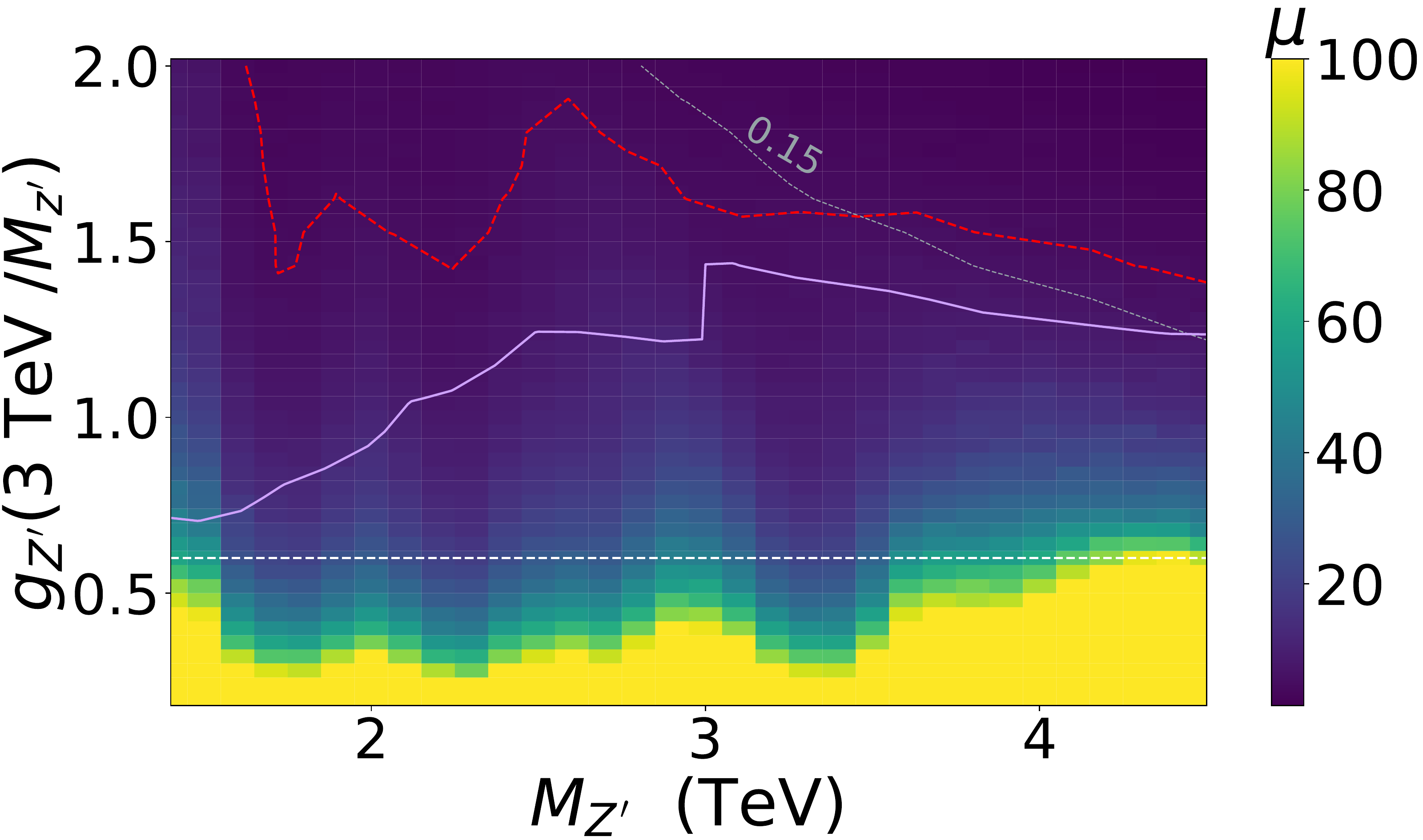}
    \caption{$\mu$ for the $2b$-tag category of the ATLAS di-jet
      search~\cite{ATLAS:2019fgd} at each point in the scanned parameter space. The current search
      does not constrain the model. The dashed red line shows our estimate of
      the HL-LHC  sensitivity. The solid purple line shows the combined
      observed bound from all LHC searches considered in this work. The dashed grey line
      shows the contour where $\Gamma /  M_{Z^\prime}$  = 0.15, and the region
      between the dashed white line and the bottom of the plot is favoured by
      the flavour and electroweak data       at the 95$\%$CL\@. Toward the extreme
      left-hand side of the plot, our predictions become more inaccurate due
      to unaccounted-for ${\mathcal O}(M_Z^2 / M_{Z^\prime}^2)$ relative corrections.}
    \label{fig:atlas_dj_2b} 
\end{figure}

\begin{figure}

    \centering
    \includegraphics[width=\columnwidth]{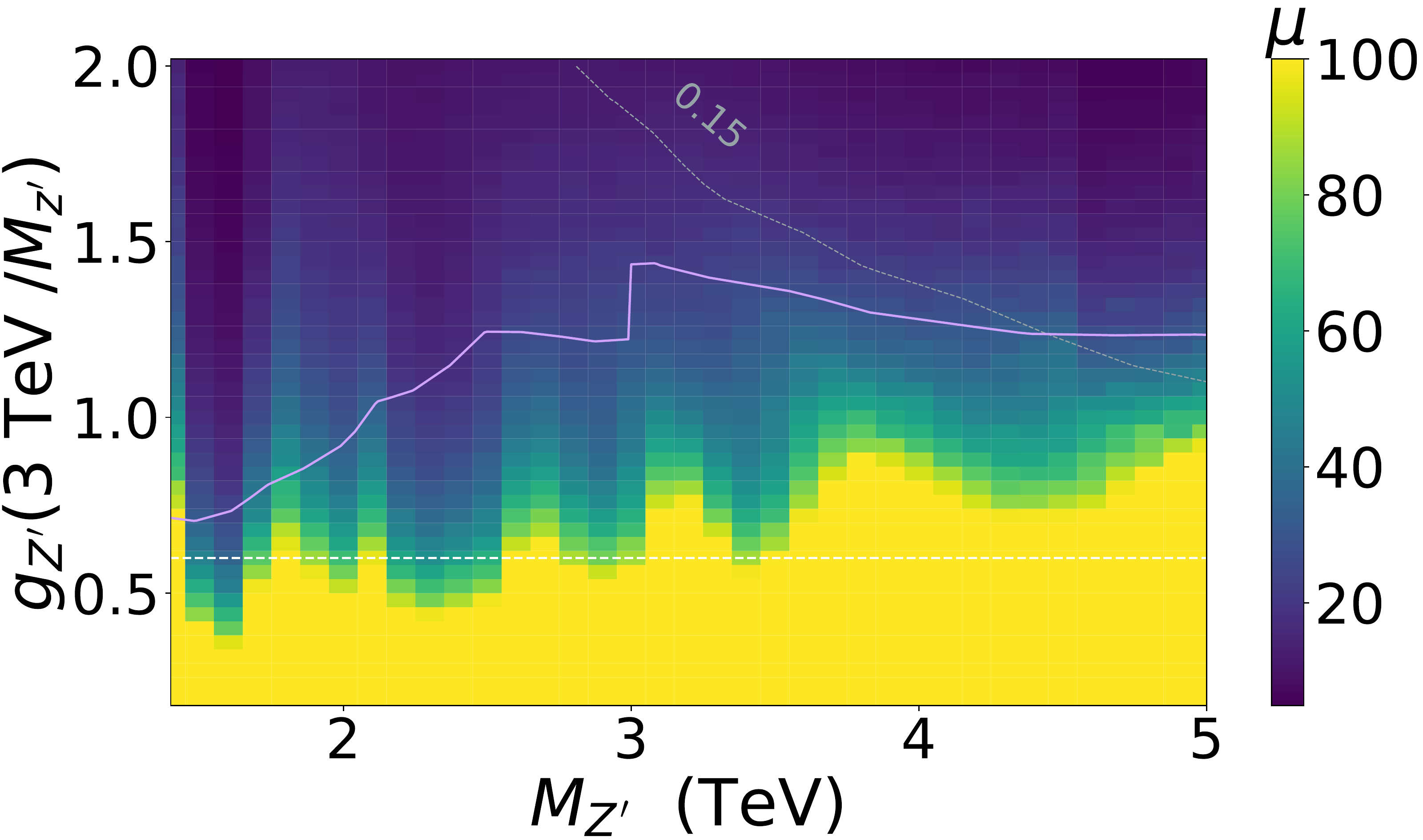}
    \caption{$\mu$ for the $1b$-tag category of the ATLAS di-jet
      search~\cite{ATLAS:2019fgd} at each point in the scanned parameter space. The current
      search does not constrain the model. The solid purple line shows the
      current observed bound from all searches considered in this work. The dashed
      grey line shows the contour where $\Gamma /  M_{Z^\prime}$  = 0.15, and
      the region between the dashed white line and the bottom of the plot is
      favoured by the flavour and electroweak data       at the 95$\%$CL\@. Toward the extreme
      left-hand side of the plot, our predictions become more inaccurate due
      to unaccounted-for ${\mathcal O}(M_Z^2 / M_{Z^\prime}^2)$ relative corrections.}
    \label{fig:atlas_dj_1b} 
\end{figure}

\begin{figure}

    \centering
    \includegraphics[width=\columnwidth]{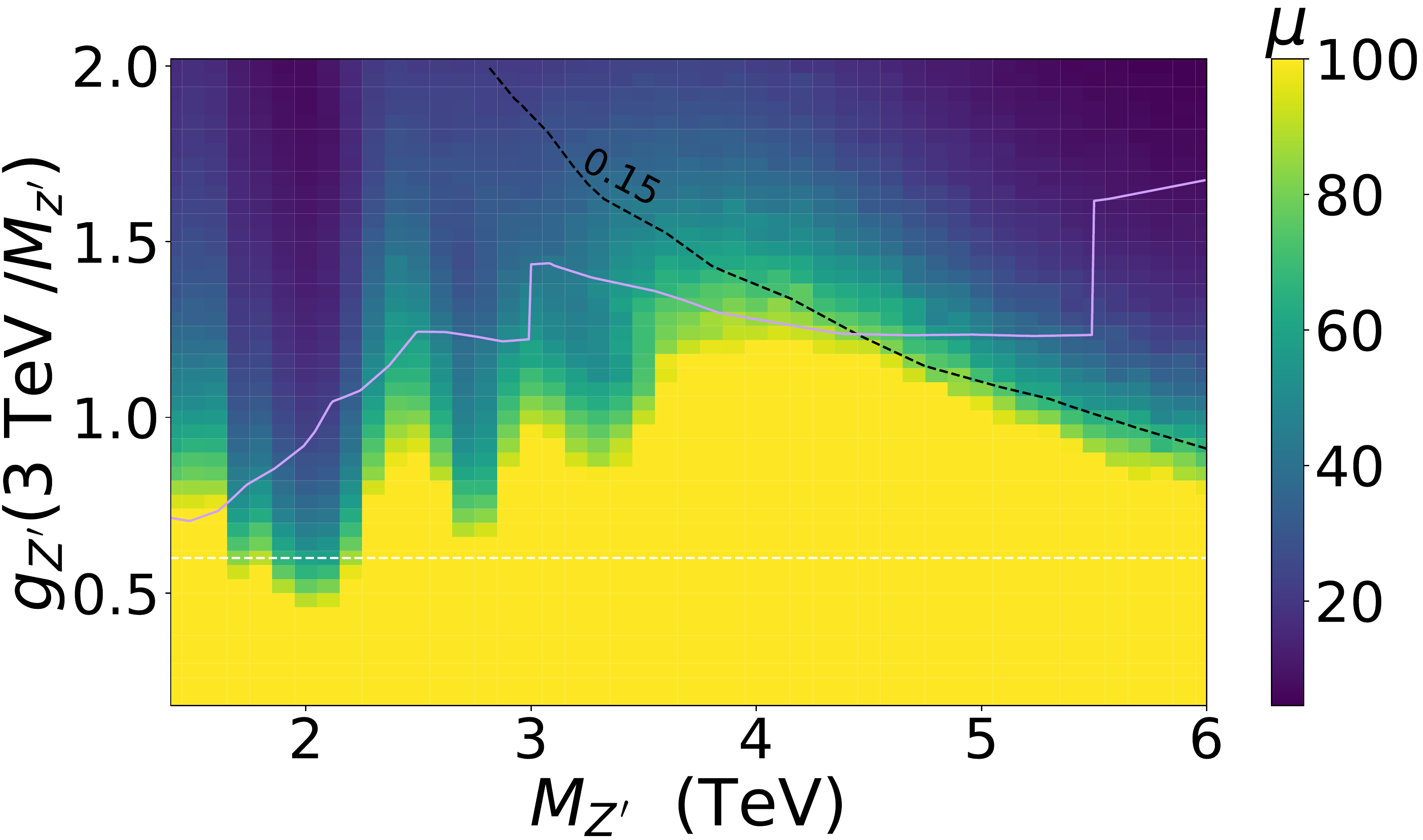}
    \caption{$\mu$ for the inclusive category of the ATLAS di-jet
      search~\cite{ATLAS:2019fgd} at each point in the scanned parameter
      space. The current search does not constrain the model. The solid
      purple line shows the current observed bound from all searches
      considered in this work. The dashed black line shows the contour where
      $\Gamma /  M_{Z^\prime}$  = 0.15, and the region between the dashed
      white line and the bottom of the plot is favoured by the flavour and
      electroweak data      at the 95$\%$CL\@. Toward the extreme
      left-hand side of the plot, our predictions become more inaccurate due
      to unaccounted-for ${\mathcal O}(M_Z^2 / M_{Z^\prime}^2)$ relative corrections.}
    \label{fig:atlas_dj_inc} 
\end{figure}

\begin{figure}

    \centering
    \includegraphics[width=\columnwidth]{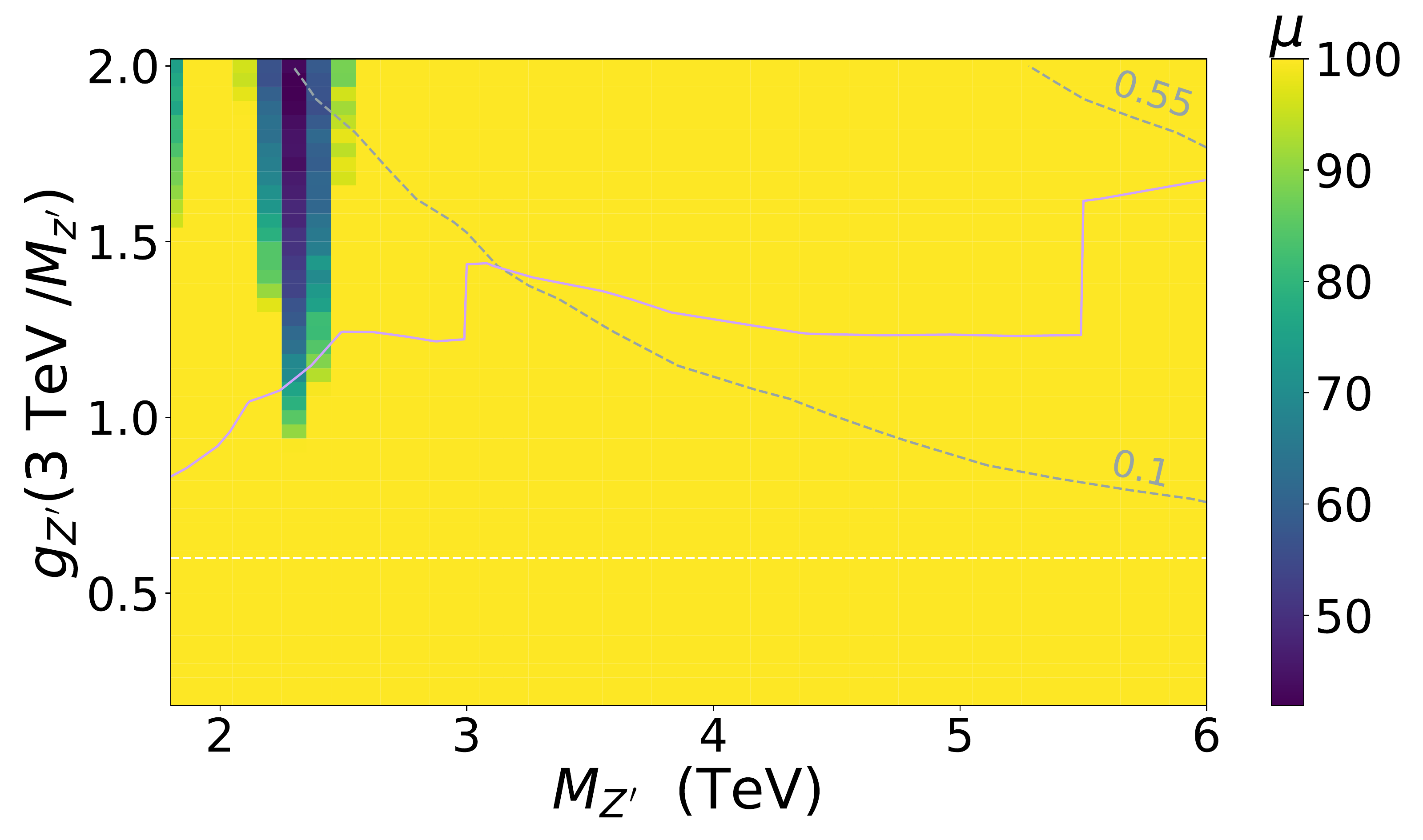}
    \caption{$\mu$ for the CMS di-jet search~\cite{CMS:2019gwf} at each point
      in the scanned parameter space. The current search does not constrain
      the model. The solid purple line shows the current observed bound from
      all searches considered in this work. The dashed grey lines show the
      contour where $\Gamma /  M_{Z^\prime}$  = 0.1 and 0.55, and the region
      between the dashed white line and the bottom of the plot is favoured by
      the flavour and electroweak data      at the 95$\%$CL\@. Toward the extreme
      left-hand side of the plot, our predictions become more inaccurate due
      to unaccounted-for ${\mathcal O}(M_Z^2 / M_{Z^\prime}^2)$ relative corrections.}
    \label{fig:cms_dj} 
\end{figure}

At present, di-jet searches leave the parameter space considered of the  TFHM
entirely unconstrained. The sensitivity of the ATLAS di-jet search is
considerably greater than that of the CMS, and increases with the number of
$b$-tagged jets. We show a plot of $\mu$ for the doubly $b$-tagged category in
Fig.~\ref{fig:atlas_dj_2b}, which we estimate would gain sensitivity to the
TFHM at the HL-LHC\@. The ATLAS di-jet singly $b$-tagged and
inclusive classes, and the CMS search remain insensitive in our forecast as
shown in Figs.~\ref{fig:atlas_dj_1b}-\ref{fig:cms_dj}, respectively.  


\begin{figure}

    \centering
    \includegraphics[width=\columnwidth]{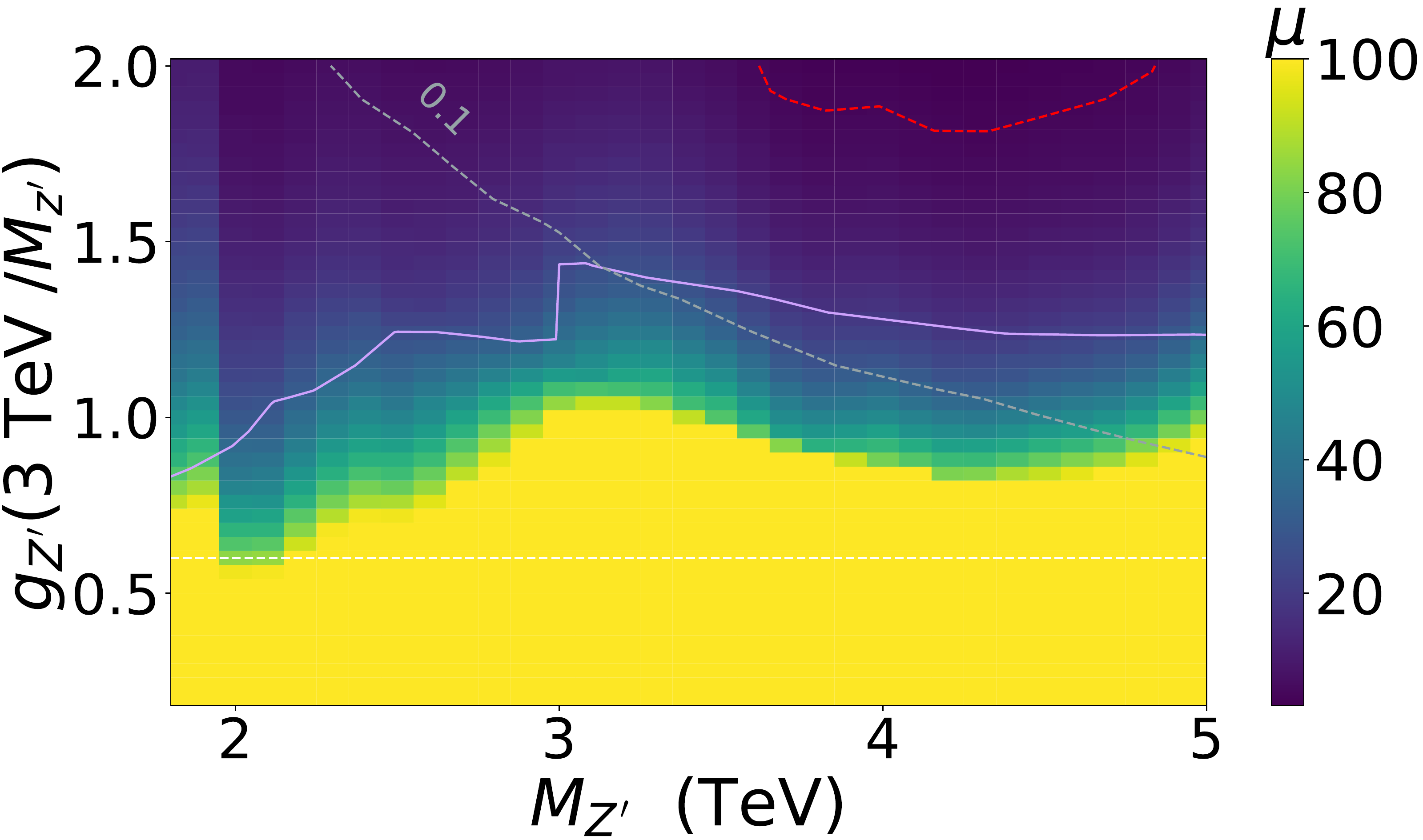}
    \caption{$\mu$ for the  ATLAS di-top search~\cite{ATLAS:2020lks} at each
      point in the scanned parameter space. The current search does not
      constrain the model. The dashed red line shows our estimate of the
      HL-LHC sensitivity. The solid purple line shows the current observed
      bound from all searches considered in this work. The dashed grey line
      shows the contour where $\Gamma /  M_{Z^\prime}$  = 0.1, and the region
      between the dashed white line and the bottom of the plot is favoured by
      the flavour and electroweak data      at the 95$\%$CL\@. Toward the extreme
      left-hand side of the plot, our predictions become more inaccurate due
      to unaccounted-for ${\mathcal O}(M_Z^2 / M_{Z^\prime}^2)$ relative corrections.}
    \label{fig:top} 
\end{figure}

The $t\bar{t}$ search results are presented for $1.9\text{~TeV}<
M_{Z^{\prime}}< \text{5~TeV}$  and does not constrain the considered parameter
space at present, as shown in Fig.~$\ref{fig:top}$. The greatest sensitivity
occurs at the upper end of this range, where the width-to-mass ratio becomes
sizeable and the model becomes non-perturbative. Our estimations (which are
based on perturbative calculations) are more unreliable in such a region of
parameter space, limiting the usefulness of this channel to the TFHM
$Z^\prime$ search.  

We have estimated the size of theoretical uncertainties on our estimated bounds in
Appendix~\ref{sec:THun}. 

\section{Discussion \label{sec:disc}}

The Third Family Hypercharge Model~\cite{Allanach:2018lvl} is a simple model
that is capable of 
explaining the \bsmm{} anomalies as well as explaining a couple of gross
features of the SM (the large masses of third family quarks and the small CKM
mixing angles). It is not intended to be the final word on the
ultra-violet limit, but rather is a next-level effective field theory beyond the SM,
valid around the $M_{Z^\prime}\sim {\mathcal O}$(TeV) scale.
The $Z^\prime$ that the model predicts, with its flavour dependent couplings,
provides an obvious target for direct searches at high energy colliders.
In the TFHM, the only sizeable coupling between the $Z^\prime$ and the quarks
is
the coupling to $b \bar b$. The
$Z^\prime$ production cross-section is much smaller in the TFHM than  in
models where for $Z^\prime$ vector bosons 
couple in a family-universal way to quarks, since in the TFHM, $Z^\prime$
production is doubly suppressed by the bottom quark parton distribution functions.
A previous computation~\cite{Allanach:2019mfl} of the bounds on TFHM parameter space
placed by the lack of a significant signal in ATLAS $Z^\prime$ direct searches 
only examined the predicted $Z^\prime \rightarrow \mu^+ \mu^-$
mode. This mode is experimentally clean, with low
backgrounds at high values of the invariant mass of the di-muon pair.

\begin{figure*}
  \centering
    \includegraphics[width=\textwidth]{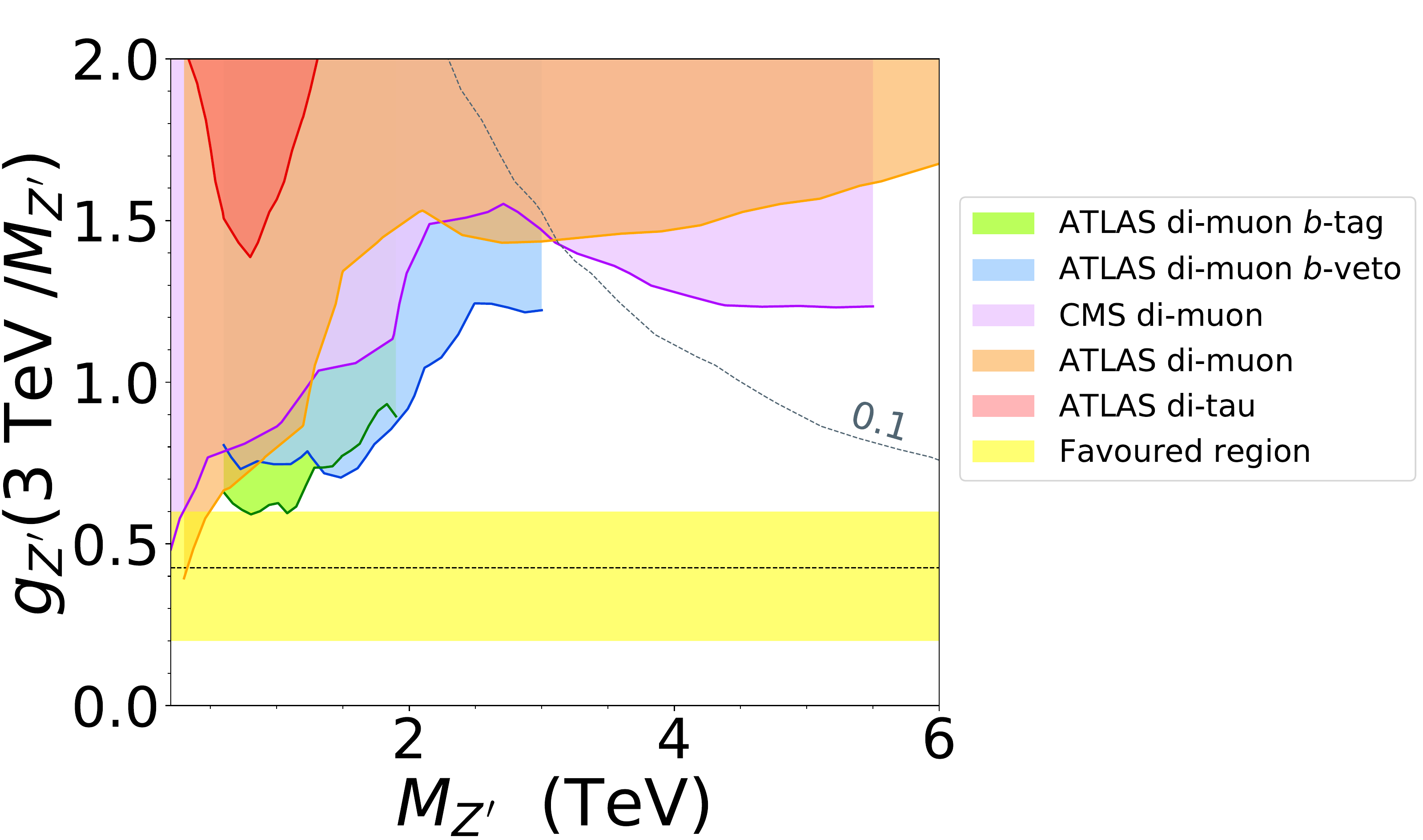}
    \caption{Summary of the 95\% CL excluded regions in the TFHM from the LHC 
      $Z^\prime$ searches 
      considered in this work. For each search, the region excluded to the
      95$\%$ CL is shown by the coloured region, as according to the legend.
      The yellow band indicates the region favoured
      by the flavour and electro-weak data       at the 95$\%$CL, and the horizontal
      dashed black line indicates the line of best-fit. The dashed grey
      line shows the contour where $\Gamma /  M_{Z^\prime}$  = 0.1. Toward the extreme
      left-hand side of the plot, our predictions become more inaccurate due
      to unaccounted-for ${\mathcal O}(M_Z^2 / M_{Z^\prime}^2)$ relative corrections.}
    \label{fig:bounds} 
\end{figure*}

Ref.~\cite{Allanach:2019mfl} found that the ATLAS di-muon search excluded
$M_{Z^\prime}>1.2$~TeV at 95$\%$ CL when the parameter space was constrained to fit the
\bsmm{} anomalies.
However, since then the \bsmm{} data
have moved in the direction of the SM predictions while remaining in tension
with them.  
A fit of the TFHM to some more recent data included a fit to
electroweak data, which further pushed the parameter space
toward the SM limit~\cite{Allanach:2021kzj}. In the present paper, we find that the
combined push toward the SM limit means smaller expected signals in di-muon
invariant mass bumps
in the TFHM, with consequently significantly weaker limits
than previously found, as shown in Fig.~\ref{fig:bounds}. In fact, the figure
shows that one can quote no lower bound on $M_{Z^\prime}$ at the 95$\%$ CL
from either ATLAS or CMS di-muon searches for parameters in the favoured
region. Our estimates of both inclusive (i.e.\ ATLAS and CMS) di-muon search 
bounds are in
approximate agreement with those in Ref.~\cite{Allanach:2021gmj}, which uses a
similar methodology but different simulation software and different parton
distribution functions.  

As Table~\ref{tab:BRs} shows, the TFHM predicts other $Z^\prime$ decay modes
that have higher branching ratios than the di-muon mode.
Our purpose here is two-fold: firstly, to compare (for the first time)
the bounds upon the model originating
from $Z^\prime$ searches in various different
channels. The parameters are arranged so that
the model fits \bsmm{} data \emph{and} electroweak data (to 95$\%$CL in the favoured
region), as obtained from
Ref.~\cite{Allanach:2021kzj}. Fig.~\ref{fig:bounds} compares current bounds
from the relevant
searches. We see that the constraints of the di-muon searches were enhanced
at larger values of $M_{Z^\prime}$ by requiring an additional $b-$tag, or
indeed by vetoing $b$ quarks. Di-tau bounds begin to constrain the high
coupling limit (but not within the favoured region of the fit). Bounds from
the non-observation of di-top or
di-jet resonances at the LHC do not yet produce bounds within the parameter
space shown 
and so they are omitted from the figure. 

One may ask the question: which of the final states considered here are likely signals in
\emph{other} $Z^\prime$ models that explain the \bsmm\ anomalies? It is obvious that
final states including di-muons must be present, since the $Z^\prime$ must
couple to them in order to mediate the \bsmm\ process. Since it must also
connect the di-muons to $b$ quarks, final states including a $b-$tag will also
necessarily be present. Di-tau final states are very common: for example they
can also been seen in two deformed TFHM models with appreciable branching
ratios~\cite{Allanach:2019iiy,Allanach:2021kzj}, however they are not strictly
necessary, as we shall now explain.
Whether the $Z^\prime$ couples to di-taus depends upon the additional-$U(1)$
charge of the third family leptons. These are constrained by 
the cancellation of quantum field theoretic
anomalies (which in turn depends upon the fermion content assumed), resulting
in model dependence.
For example, in the 
$U(1)_{B_3-L_2}$ model~\cite{Alonso:2017uky,Bonilla:2017lsq,Allanach:2020kss}, the quantum field theoretic
anomalies are cancelled by a heavy right-handed muon neutrino and there is no
$Z^\prime$ coupling to di-tau final states.
Since the $Z^\prime$ must couple to \emph{left-handed}
bottom quarks to successfully fit \bsmm\ data, a coupling to tops is
also guaranteed, since the left-handed top is unified with the left-handed
bottom within an $SU(2)_L$ doublet. It almost goes without saying that the
numerical values of $Z^\prime$ branching ratios into the possible final states are
dependent on additional-$U(1)$ charges and are therefore model dependent.  As found for the TFHM model in this study, the strongest constraints on other $Z^\prime$ models are also likely to come from the di-muon channel, both with and without $b$-tagging.

\begin{figure*}
  \centering
    \includegraphics[width=\textwidth]{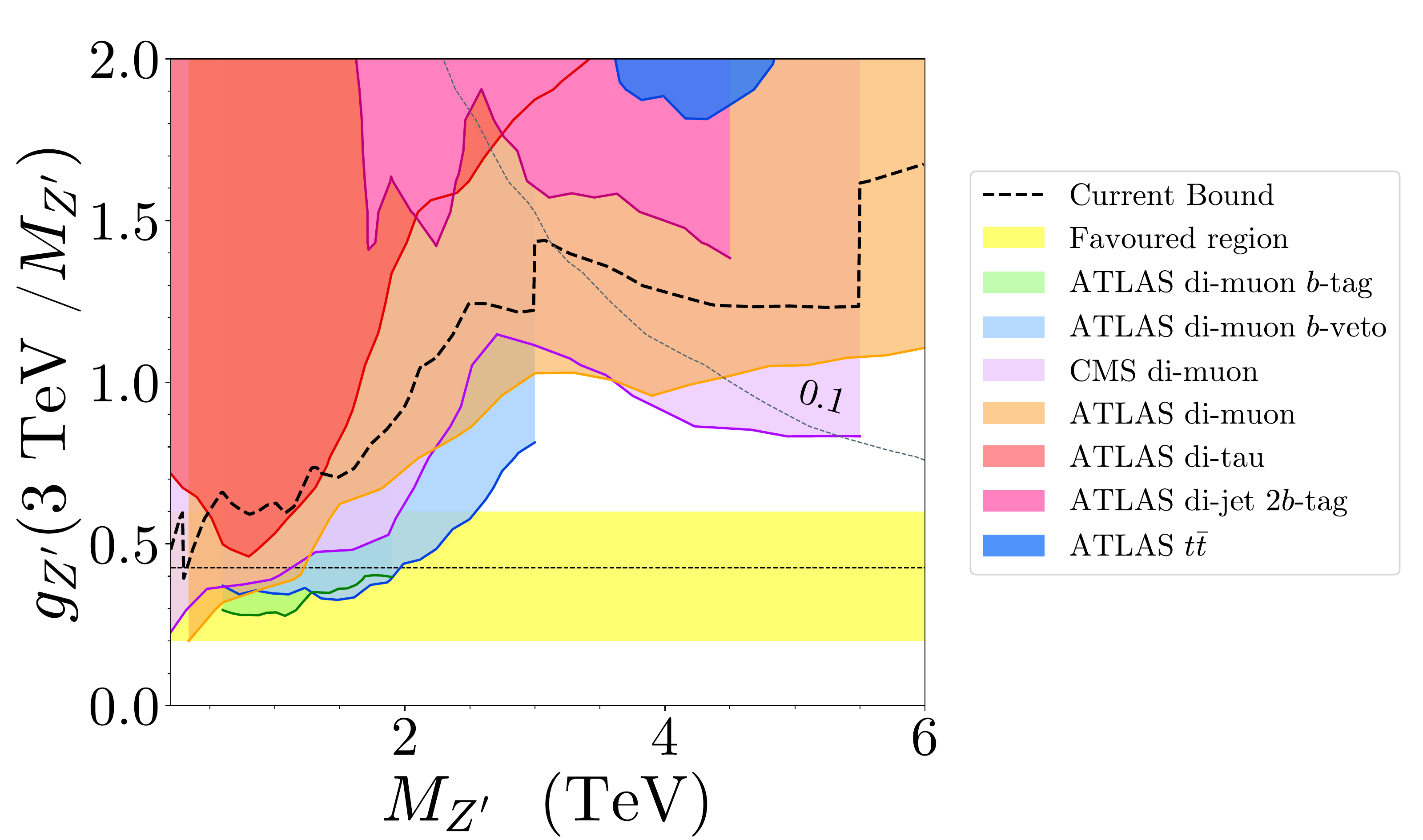}
    \caption{Summary of our estimate of the 95\% CL sensitivity regions of
      HL-LHC $Z^\prime$ searches in the TFHM\@. The yellow band indicates the
      region favoured 
      by the flavour and electro-weak data       at the 95$\%$CL, and the horizontal
      dashed black line indicates the line of best fit. The region
      above the dashed grey
      line has $\Gamma /  M_{Z^\prime} > 0.1$. Toward the extreme
      left-hand side of the plot, our predictions become more inaccurate due
      to unaccounted-for ${\mathcal O}(M_Z^2 / M_{Z^\prime}^2)$ relative corrections.}
    \label{fig:sensitivities} 
\end{figure*}
The second purpose of our paper
is to provide a rough estimate of the HL-LHC sensitivity
in each channel, looking forward to the prospects of potential discovery of
the $Z^\prime$ and tests of the model by measuring various different decay modes. 
We summarise this in Fig.~\ref{fig:sensitivities}. 
The figure shows that, while the currently favoured region is barely being
constrained by current LHC searches, the HL-LHC has sensitivity to an
appreciable part of parameter space in the region $M_{Z^\prime} < 2.5$~TeV in various
modes involving di-muons, with or without $b-$tags and $b-$vetoes. There is the
possibility within the favoured region of joint sensitivity between the
various di-muon
searches and the ATLAS di-tau search at around $M_{Z^\prime}=1$~TeV.
As such, we look forward to the HL-LHC run, which will extend the sensitivity
of ATLAS and CMS to the TFHM (and to other models of its ilk).

\section*{Acknowledgements}
This work has been partially supported by STFC Consolidated HEP grants
ST/P000681/1 and ST/T000694/1. We thank other members of the
Cambridge Pheno Working Group for helpful discussions, especially B Webber. 
We also thank J Butterworth for detailed checks upon the $Z^\prime$ LHC production
cross-section and the resulting inclusive di-muon bounds.

\appendix

\section{Theoretical Uncertainties \label{sec:THun}}

 It is important to note that our simulated event samples are subject to
 several sources of theoretical uncertainty. We quantify the impact of theory
 uncertainties on the location of our exclusion bounds, by finding the
 exclusion limits from the upper and lower estimates of the observable due to
 uncertainty. We estimate the dominant systematic contributions  on our
 production cross-section computations using the ``systematics" program
 included in {\tt Madgraph}. This estimates QCD scale uncertainties by varying
 the factorisation and renormalisation scales simultaneously up and down by a
 factor of 2. The emission scale uncertainty is assessed independently by an
 analogous procedure. An uncertainty due to the choice of PDF is obtained by
 from the error set replicas. We add the scale, emission and PDF errors
 linearly to obtain an estimate of the total theoretical uncertainty on the
 production cross-section. Statistical uncertainties arising from the
 Monte-Carlo integration procedure are found to be significantly smaller in
 magnitude and we have neglected them in this evaluation.

\begin{figure}

    \centering
    \includegraphics[width=\columnwidth]{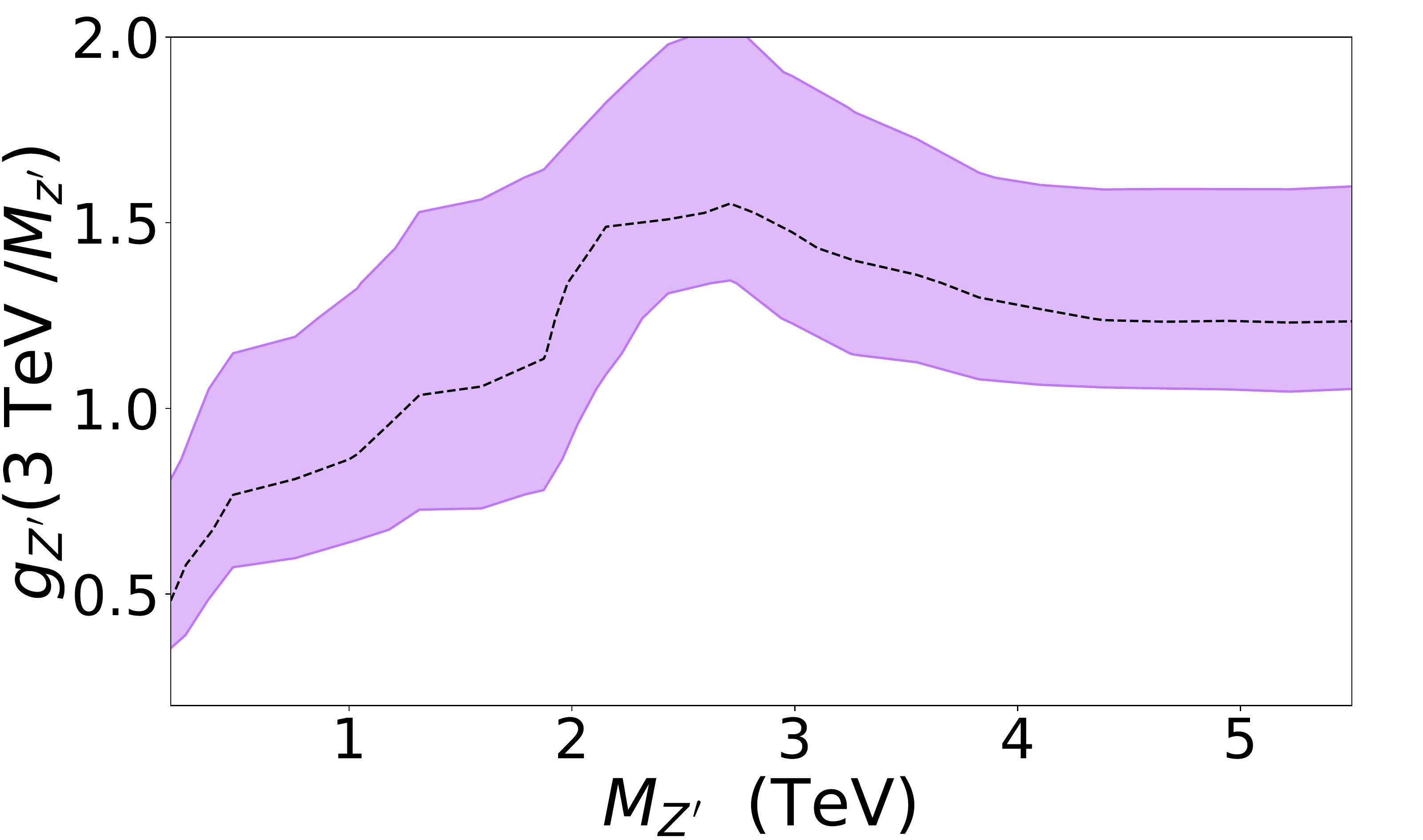}
    \caption{Effects of theoretical uncertainties on bounds from the CMS di-muon search~\cite{CMS:2021ctt}.The dashed black line shows the central limit. The purple region falls between bounds obtained from upper and lower estimates of the cross-section accounting for systematic uncertainties. Toward the extreme
      left-hand side of the plot, our predictions become more inaccurate due
      to unaccounted-for ${\mathcal O}(M_Z^2 / M_{Z^\prime}^2)$ relative corrections.}
    \label{fig:cms_dm_er} 
\end{figure}

\begin{figure}

    \centering
    \includegraphics[width=\columnwidth]{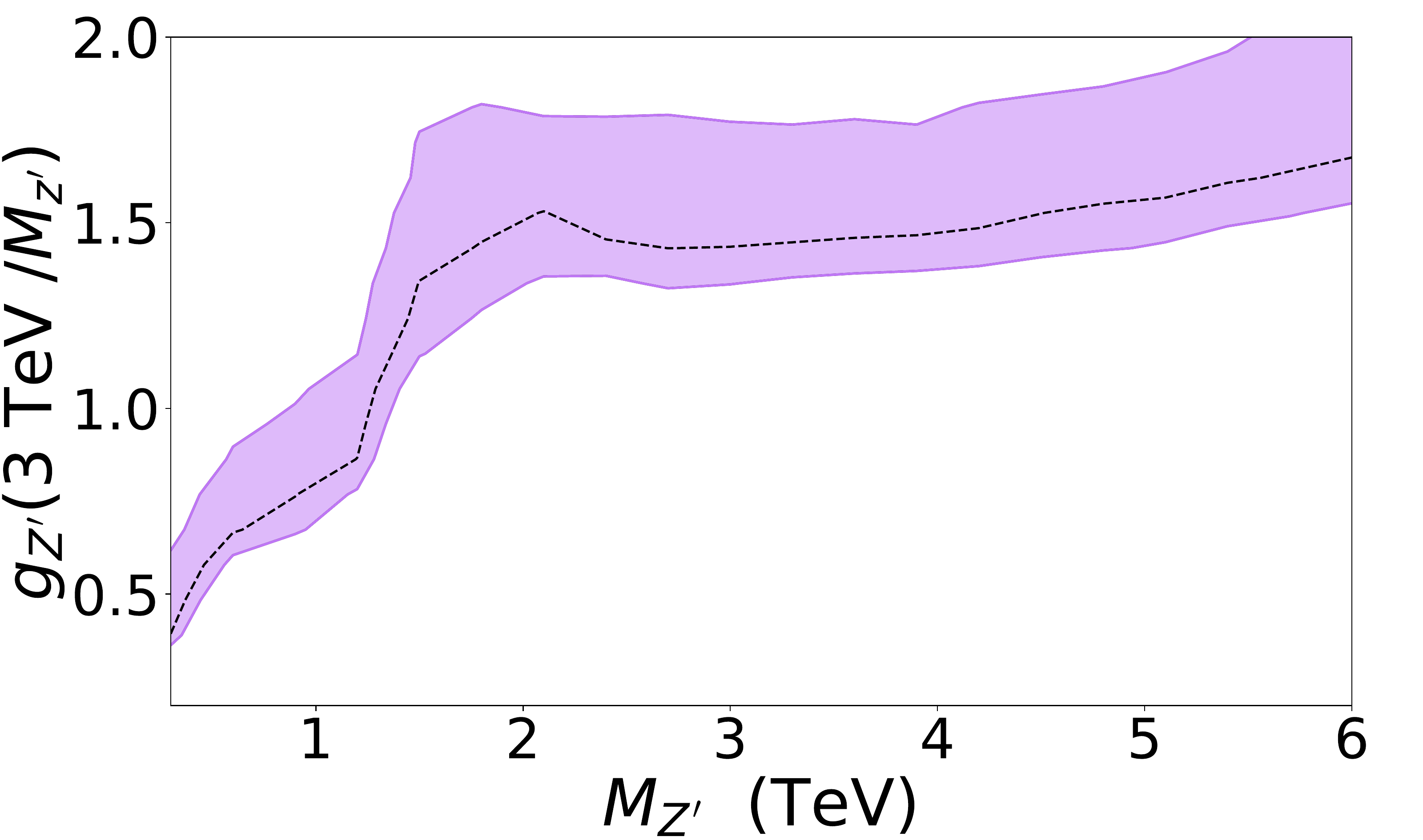}
    \caption{Effects of theoretical uncertainties on bounds from the ATLAS di-muon search~\cite{ATLAS:2019erb}.The dashed black line shows the central limit. The purple region falls between bounds obtained from upper and lower estimates of the cross-section accounting for systematic uncertainties. Toward the extreme
      left-hand side of the plot, our predictions become more inaccurate due
      to unaccounted-for ${\mathcal O}(M_Z^2 / M_{Z^\prime}^2)$ relative corrections.}
    \label{fig:atlas_dm_er}
\end{figure}

\begin{figure}

    \centering
    \includegraphics[width=\columnwidth]{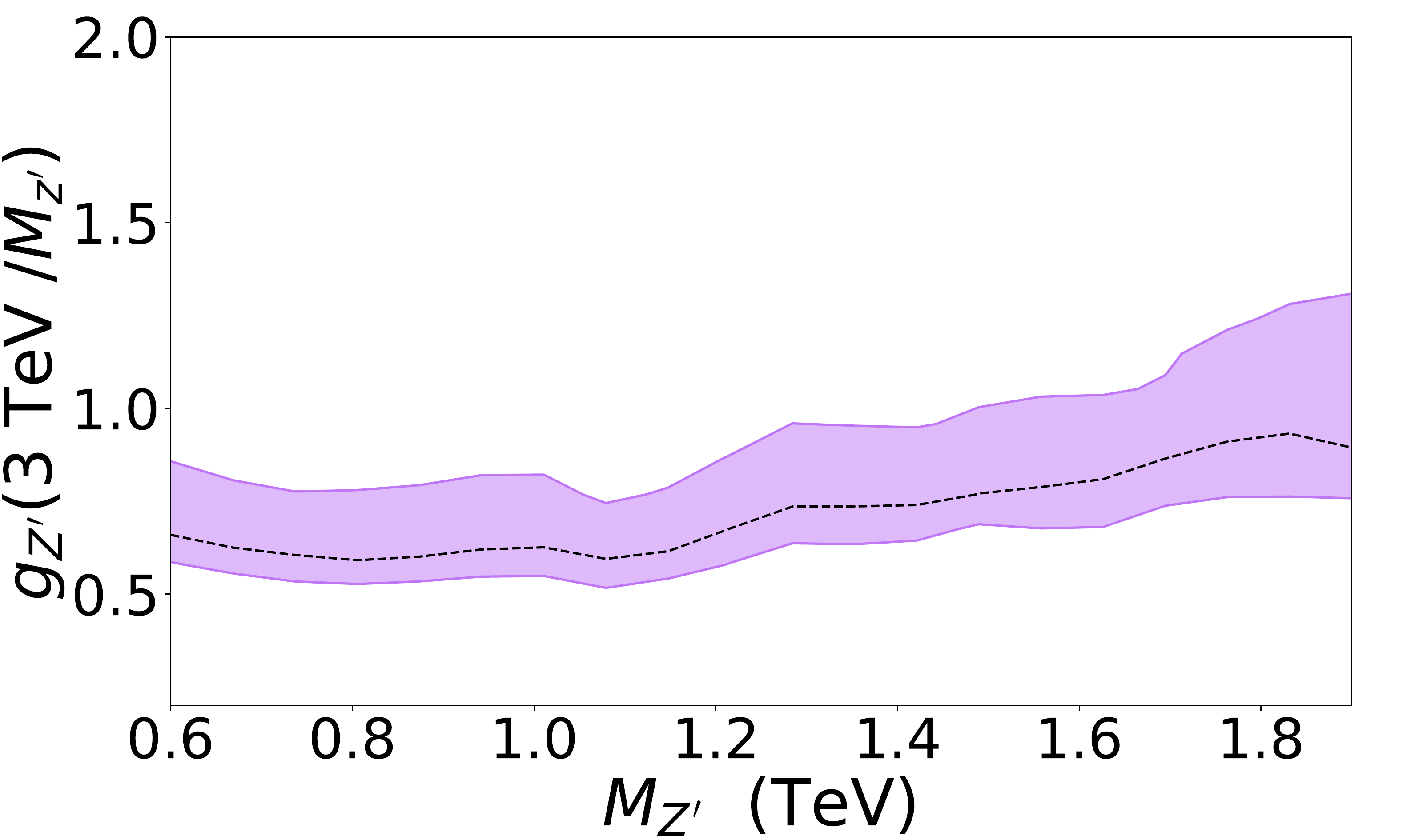}
    \caption{Effects of theoretical uncertainties on bunds from the $b$-tag category of the ATLAS di-muon search with $b$-tagging~\cite{ATLAS:2021mla}.The dashed black line shows the central limit. The purple region falls between bounds obtained from upper and lower estimates of the cross-section accounting for systematic uncertainties. Toward the extreme
      left-hand side of the plot, our predictions become more inaccurate due
      to unaccounted-for ${\mathcal O}(M_Z^2 / M_{Z^\prime}^2)$ relative corrections.}
    \label{fig:atlas_dm_btag_er} 
\end{figure}

\begin{figure}

    \centering
    \includegraphics[width=\columnwidth]{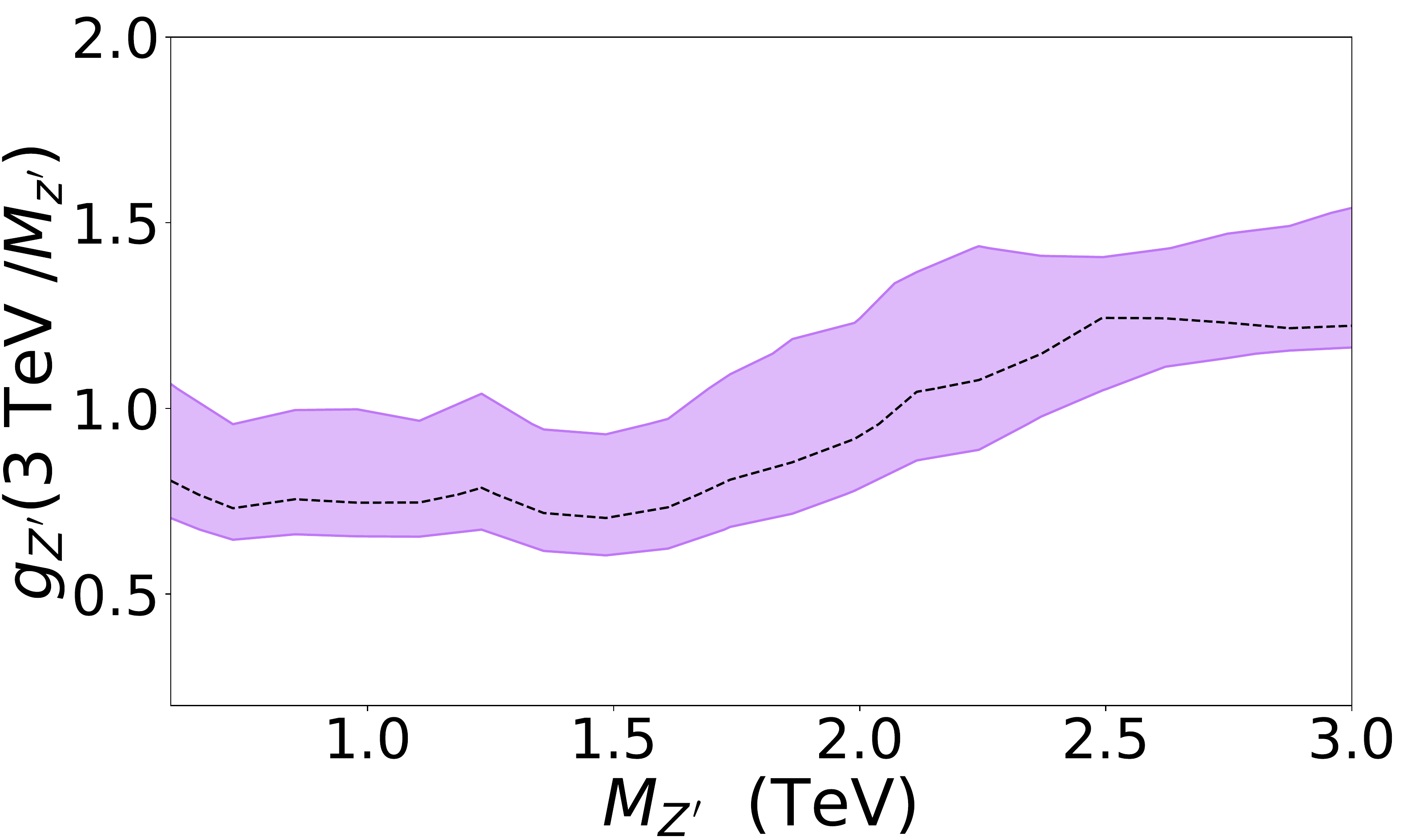}
    \caption{Effects of theoretical uncertainties on bunds from the $b$-veto category of the ATLAS di-muon search with $b$-tagging~\cite{ATLAS:2021mla}.The dashed black line shows the central limit. The purple region falls between bounds obtained from upper and lower estimates of the cross-section accounting for systematic uncertainties. Toward the extreme
      left-hand side of the plot, our predictions become more inaccurate due
      to unaccounted-for ${\mathcal O}(M_Z^2 / M_{Z^\prime}^2)$ relative corrections.}
    \label{fig:atlas_dm_bveto_er} 
\end{figure}

\begin{figure}

    \centering
    \includegraphics[width=\columnwidth]{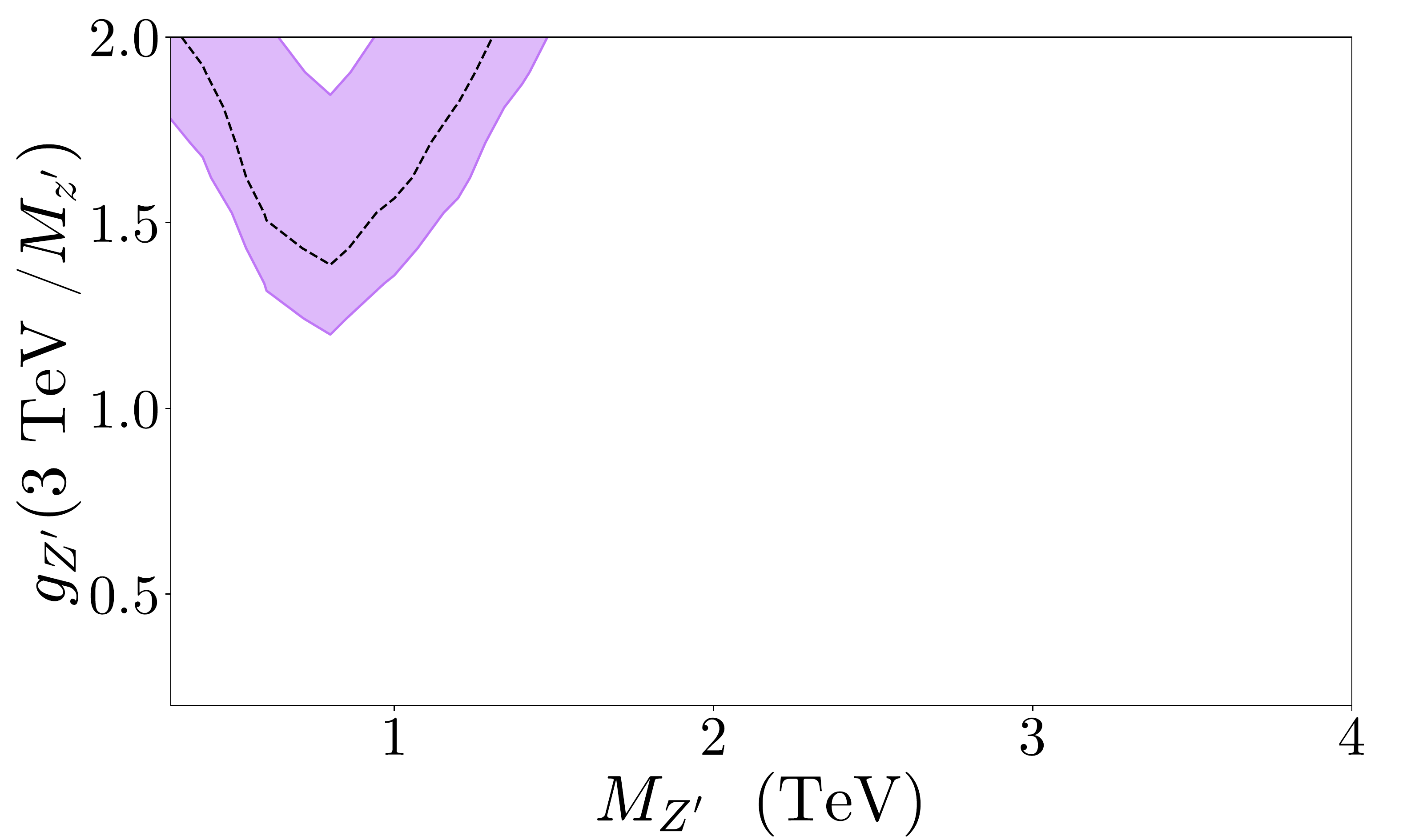}
    \caption{Effects of theoretical uncertainties on bounds from the ATLAS di-tau search~\cite{Aaboud_2018}.The dashed black line shows the central limit. The purple region falls between bounds obtained from upper and lower estimates of the cross-section accounting for systematic uncertainties. Toward the extreme
      left-hand side of the plot, our predictions become more inaccurate due
      to unaccounted-for ${\mathcal O}(M_Z^2 / M_{Z^\prime}^2)$ relative corrections.}
    \label{fig:tau_errors} 
\end{figure}

Figs.~\ref{fig:cms_dm_er}-\ref{fig:tau_errors} show the impact of theoretical uncertainties on the calculated bounds, of the CMS di-muon, ATLAS di-muon, ATLAS di-muon with $b$-tag and $b$-veto, and ATLAS di-tau searches respectively. 
In all cases, the theoretical uncertainties are appreciable. They
could be reduced in the future by calculating to an order of perturbation
theory beyond the tree level.

\bibliographystyle{JHEP-2}
\bibliography{search}

\providecommand{\href}[2]{#2}\begingroup\raggedright\begin{thebibliography}{10}

\bibitem{Allanach:2018lvl}
B.~C. Allanach and J.~Davighi, {\it {Third family hypercharge model for $
  {R}_{K^{\left(\ast \right)}} $ and aspects of the fermion mass problem}},
  {\em JHEP} {\bf 12} (2018) 075 [\href{http://arXiv.org/abs/1809.01158}{{\tt
  1809.01158}}].

\bibitem{Aaij:2021vac}
{\bf LHCb} Collaboration, R.~Aaij {\em et.~al.}, {\it {Test of lepton
  universality in beauty-quark decays}},
  \href{http://arXiv.org/abs/2103.11769}{{\tt 2103.11769}}.

\bibitem{Aaij:2013qta}
{\bf LHCb} Collaboration, R.~Aaij {\em et.~al.}, {\it {Measurement of
  Form-Factor-Independent Observables in the Decay $B^{0} \to K^{*0} \mu^+
  \mu^-$}},  {\em Phys. Rev. Lett.} {\bf 111} (2013) 191801
  [\href{http://arXiv.org/abs/1308.1707}{{\tt 1308.1707}}].

\bibitem{Aaij:2015oid}
{\bf LHCb} Collaboration, R.~Aaij {\em et.~al.}, {\it {Angular analysis of the
  $B^{0} \to K^{*0} \mu^{+} \mu^{-}$ decay using 3 fb$^{-1}$ of integrated
  luminosity}},  {\em JHEP} {\bf 02} (2016) 104
  [\href{http://arXiv.org/abs/1512.04442}{{\tt 1512.04442}}].

\bibitem{Aaboud:2018krd}
{\bf ATLAS} Collaboration, M.~Aaboud {\em et.~al.}, {\it {Angular analysis of
  $B^0_d \rightarrow K^{*}\mu^+\mu^-$ decays in $pp$ collisions at $\sqrt{s}=
  8$ TeV with the ATLAS detector}},  {\em JHEP} {\bf 10} (2018) 047
  [\href{http://arXiv.org/abs/1805.04000}{{\tt 1805.04000}}].

\bibitem{Sirunyan:2017dhj}
{\bf CMS} Collaboration, A.~M. Sirunyan {\em et.~al.}, {\it {Measurement of
  angular parameters from the decay $\mathrm{B}^0 \to \mathrm{K}^{*0} \mu^+
  \mu^-$ in proton-proton collisions at $\sqrt{s} = $ 8 TeV}},  {\em Phys.
  Lett. B} {\bf 781} (2018) 517--541
  [\href{http://arXiv.org/abs/1710.02846}{{\tt 1710.02846}}].

\bibitem{Khachatryan:2015isa}
{\bf CMS} Collaboration, V.~Khachatryan {\em et.~al.}, {\it {Angular analysis
  of the decay $B^0 \to K^{*0} \mu^+ \mu^-$ from pp collisions at $\sqrt s = 8$
  TeV}},  {\em Phys. Lett. B} {\bf 753} (2016) 424--448
  [\href{http://arXiv.org/abs/1507.08126}{{\tt 1507.08126}}].

\bibitem{Bobeth:2017vxj}
C.~Bobeth, M.~Chrzaszcz, D.~van Dyk and J.~Virto, {\it {Long-distance effects
  in $B\rightarrow K^*\ell \ell $ from analyticity}},  {\em Eur. Phys. J. C}
  {\bf 78} (2018), no.~6 451 [\href{http://arXiv.org/abs/1707.07305}{{\tt
  1707.07305}}].

\bibitem{Aaboud:2018mst}
{\bf ATLAS} Collaboration, M.~Aaboud {\em et.~al.}, {\it {Study of the rare
  decays of $B^0_s$ and $B^0$ mesons into muon pairs using data collected
  during 2015 and 2016 with the ATLAS detector}},  {\em JHEP} {\bf 04} (2019)
  098 [\href{http://arXiv.org/abs/1812.03017}{{\tt 1812.03017}}].

\bibitem{Chatrchyan:2013bka}
{\bf CMS} Collaboration, S.~Chatrchyan {\em et.~al.}, {\it {Measurement of the
  $B^0_s \to \mu^+ \mu^-$ Branching Fraction and Search for $B^0 \to \mu^+
  \mu^-$ with the CMS Experiment}},  {\em Phys. Rev. Lett.} {\bf 111} (2013)
  101804 [\href{http://arXiv.org/abs/1307.5025}{{\tt 1307.5025}}].

\bibitem{CMS:2014xfa}
{\bf CMS, LHCb} Collaboration, V.~Khachatryan {\em et.~al.}, {\it {Observation
  of the rare $B^0_s\to\mu^+\mu^-$ decay from the combined analysis of CMS and
  LHCb data}},  {\em Nature} {\bf 522} (2015) 68--72
  [\href{http://arXiv.org/abs/1411.4413}{{\tt 1411.4413}}].

\bibitem{Aaij:2017vad}
{\bf LHCb} Collaboration, R.~Aaij {\em et.~al.}, {\it {Measurement of the
  $B^0_s\to\mu^+\mu^-$ branching fraction and effective lifetime and search for
  $B^0\to\mu^+\mu^-$ decays}},  {\em Phys. Rev. Lett.} {\bf 118} (2017), no.~19
  191801 [\href{http://arXiv.org/abs/1703.05747}{{\tt 1703.05747}}].

\bibitem{LHCbtalk}
{\bf LHCb} Collaboration, K.~Petridis and M.~Santimaria, {\it New results on
  theoretically clean observables in rare b-meson decays from lhcb},  3, 2021.
\newblock LHC seminar.

\bibitem{Aaij:2015esa}
{\bf LHCb} Collaboration, R.~Aaij {\em et.~al.}, {\it {Angular analysis and
  differential branching fraction of the decay $B^0_s\to\phi\mu^+\mu^-$}},
  {\em JHEP} {\bf 09} (2015) 179 [\href{http://arXiv.org/abs/1506.08777}{{\tt
  1506.08777}}].

\bibitem{CDF:2012qwd}
{CDF collaboration}, {\it {Precise Measurements of Exclusive b
  \textrightarrow{} s\textmu{}+\textmu{} \ensuremath{-} Decay Amplitudes Using
  the Full CDF Data Set}},  {\em CDF-NOTE-10894} (6, 2012).

\bibitem{Lancierini:2021sdf}
D.~Lancierini, G.~Isidori, P.~Owen and N.~Serra, {\it {On the significance of
  new physics in $b\to s\ell^+\ell^-$ decays}},
  \href{http://arXiv.org/abs/2104.05631}{{\tt 2104.05631}}.

\bibitem{Altmannshofer:2014cfa}
W.~Altmannshofer, S.~Gori, M.~Pospelov and I.~Yavin, {\it {Quark flavor
  transitions in $L_\mu-L_\tau$ models}},  {\em Phys. Rev. D} {\bf 89} (2014)
  095033 [\href{http://arXiv.org/abs/1403.1269}{{\tt 1403.1269}}].

\bibitem{Crivellin:2015mga}
A.~Crivellin, G.~D'Ambrosio and J.~Heeck, {\it {Explaining
  $h\to\mu^\pm\tau^\mp$, $B\to K^* \mu^+\mu^-$ and $B\to K \mu^+\mu^-/B\to K
  e^+e^-$ in a two-Higgs-doublet model with gauged $L_\mu-L_\tau$}},  {\em
  Phys. Rev. Lett.} {\bf 114} (2015) 151801
  [\href{http://arXiv.org/abs/1501.00993}{{\tt 1501.00993}}].

\bibitem{Crivellin:2015lwa}
A.~Crivellin, G.~D'Ambrosio and J.~Heeck, {\it {Addressing the LHC flavor
  anomalies with horizontal gauge symmetries}},  {\em Phys. Rev. D} {\bf 91}
  (2015), no.~7 075006 [\href{http://arXiv.org/abs/1503.03477}{{\tt
  1503.03477}}].

\bibitem{Crivellin:2015era}
A.~Crivellin, L.~Hofer, J.~Matias, U.~Nierste, S.~Pokorski and J.~Rosiek, {\it
  {Lepton-flavour violating $B$ decays in generic $Z'$ models}},  {\em Phys.
  Rev. D} {\bf 92} (2015), no.~5 054013
  [\href{http://arXiv.org/abs/1504.07928}{{\tt 1504.07928}}].

\bibitem{Altmannshofer:2015mqa}
W.~Altmannshofer and I.~Yavin, {\it {Predictions for lepton flavor universality
  violation in rare B decays in models with gauged $L_\mu - L_\tau$}},  {\em
  Phys. Rev. D} {\bf 92} (2015), no.~7 075022
  [\href{http://arXiv.org/abs/1508.07009}{{\tt 1508.07009}}].

\bibitem{Sierra:2015fma}
D.~Aristizabal~Sierra, F.~Staub and A.~Vicente, {\it {Shedding light on the
  $b\to s$ anomalies with a dark sector}},  {\em Phys. Rev. D} {\bf 92} (2015),
  no.~1 015001 [\href{http://arXiv.org/abs/1503.06077}{{\tt 1503.06077}}].

\bibitem{Celis:2015ara}
A.~Celis, J.~Fuentes-Martin, M.~Jung and H.~Serodio, {\it {Family nonuniversal
  Z' models with protected flavor-changing interactions}},  {\em Phys. Rev. D}
  {\bf 92} (2015), no.~1 015007 [\href{http://arXiv.org/abs/1505.03079}{{\tt
  1505.03079}}].

\bibitem{Greljo:2015mma}
A.~Greljo, G.~Isidori and D.~Marzocca, {\it {On the breaking of Lepton Flavor
  Universality in B decays}},  {\em JHEP} {\bf 07} (2015) 142
  [\href{http://arXiv.org/abs/1506.01705}{{\tt 1506.01705}}].

\bibitem{Falkowski:2015zwa}
A.~Falkowski, M.~Nardecchia and R.~Ziegler, {\it {Lepton Flavor
  Non-Universality in B-meson Decays from a U(2) Flavor Model}},  {\em JHEP}
  {\bf 11} (2015) 173 [\href{http://arXiv.org/abs/1509.01249}{{\tt
  1509.01249}}].

\bibitem{Chiang:2016qov}
C.-W. Chiang, X.-G. He and G.~Valencia, {\it {Z' model for $b\rightarrow{}sll$
  flavor anomalies}},  {\em Phys. Rev. D} {\bf 93} (2016), no.~7 074003
  [\href{http://arXiv.org/abs/1601.07328}{{\tt 1601.07328}}].

\bibitem{Boucenna:2016wpr}
S.~M. Boucenna, A.~Celis, J.~Fuentes-Martin, A.~Vicente and J.~Virto, {\it
  {Non-abelian gauge extensions for B-decay anomalies}},  {\em Phys. Lett. B}
  {\bf 760} (2016) 214--219 [\href{http://arXiv.org/abs/1604.03088}{{\tt
  1604.03088}}].

\bibitem{Boucenna:2016qad}
S.~M. Boucenna, A.~Celis, J.~Fuentes-Martin, A.~Vicente and J.~Virto, {\it
  {Phenomenology of an $SU(2) \times SU(2) \times U(1)$ model with
  lepton-flavour non-universality}},  {\em JHEP} {\bf 12} (2016) 059
  [\href{http://arXiv.org/abs/1608.01349}{{\tt 1608.01349}}].

\bibitem{Ko:2017lzd}
P.~Ko, Y.~Omura, Y.~Shigekami and C.~Yu, {\it {LHCb anomaly and B physics in
  flavored Z' models with flavored Higgs doublets}},  {\em Phys. Rev. D} {\bf
  95} (2017), no.~11 115040 [\href{http://arXiv.org/abs/1702.08666}{{\tt
  1702.08666}}].

\bibitem{Alonso:2017bff}
R.~Alonso, P.~Cox, C.~Han and T.~T. Yanagida, {\it {Anomaly-free local
  horizontal symmetry and anomaly-full rare B-decays}},  {\em Phys. Rev. D}
  {\bf 96} (2017), no.~7 071701 [\href{http://arXiv.org/abs/1704.08158}{{\tt
  1704.08158}}].

\bibitem{Tang:2017gkz}
Y.~Tang and Y.-L. Wu, {\it {Flavor non-universal gauge interactions and
  anomalies in B-meson decays}},  {\em Chin. Phys. C} {\bf 42} (2018), no.~3
  033104 [\href{http://arXiv.org/abs/1705.05643}{{\tt 1705.05643}}]. [Erratum:
  Chin.Phys.C 44, 069101 (2020)].

\bibitem{Bhatia:2017tgo}
D.~Bhatia, S.~Chakraborty and A.~Dighe, {\it {Neutrino mixing and $R_K$ anomaly
  in U(1)$_X$ models: a bottom-up approach}},  {\em JHEP} {\bf 03} (2017) 117
  [\href{http://arXiv.org/abs/1701.05825}{{\tt 1701.05825}}].

\bibitem{Fuyuto:2017sys}
K.~Fuyuto, H.-L. Li and J.-H. Yu, {\it {Implications of hidden gauged $U(1)$
  model for $B$ anomalies}},  {\em Phys. Rev. D} {\bf 97} (2018), no.~11 115003
  [\href{http://arXiv.org/abs/1712.06736}{{\tt 1712.06736}}].

\bibitem{Bian:2017xzg}
L.~Bian, H.~M. Lee and C.~B. Park, {\it {$B$-meson anomalies and Higgs physics
  in flavored $U(1)'$ model}},  {\em Eur. Phys. J. C} {\bf 78} (2018), no.~4
  306 [\href{http://arXiv.org/abs/1711.08930}{{\tt 1711.08930}}].

\bibitem{Alonso:2017uky}
R.~Alonso, P.~Cox, C.~Han and T.~T. Yanagida, {\it {Flavoured $B-L$ local
  symmetry and anomalous rare $B$ decays}},  {\em Phys. Lett. B} {\bf 774}
  (2017) 643--648 [\href{http://arXiv.org/abs/1705.03858}{{\tt 1705.03858}}].

\bibitem{Bonilla:2017lsq}
C.~Bonilla, T.~Modak, R.~Srivastava and J.~W.~F. Valle, {\it
  {$U(1)_{B_3-3L_\mu}$ gauge symmetry as a simple description of $b\to s$
  anomalies}},  {\em Phys. Rev. D} {\bf 98} (2018), no.~9 095002
  [\href{http://arXiv.org/abs/1705.00915}{{\tt 1705.00915}}].

\bibitem{King:2018fcg}
S.~F. King, {\it {$ {R}_{K^{\left(*\right)}} $ and the origin of Yukawa
  couplings}},  {\em JHEP} {\bf 09} (2018) 069
  [\href{http://arXiv.org/abs/1806.06780}{{\tt 1806.06780}}].

\bibitem{Duan:2018akc}
G.~H. Duan, X.~Fan, M.~Frank, C.~Han and J.~M. Yang, {\it {A minimal
  $U(1)^\prime$ extension of MSSM in light of the B decay anomaly}},  {\em
  Phys. Lett. B} {\bf 789} (2019) 54--58
  [\href{http://arXiv.org/abs/1808.04116}{{\tt 1808.04116}}].

\bibitem{Kang:2019vng}
Z.~Kang and Y.~Shigekami, {\it {$(g-2)_{\mu}$ versus flavor changing neutral
  current induced by the light $(B-L)_{\mu\tau}$ boson}},  {\em JHEP} {\bf 11}
  (2019) 049 [\href{http://arXiv.org/abs/1905.11018}{{\tt 1905.11018}}].

\bibitem{Calibbi:2019lvs}
L.~Calibbi, A.~Crivellin, F.~Kirk, C.~A. Manzari and L.~Vernazza, {\it
  {$Z^\prime$ models with less-minimal flavour violation}},  {\em Phys. Rev. D}
  {\bf 101} (2020), no.~9 095003 [\href{http://arXiv.org/abs/1910.00014}{{\tt
  1910.00014}}].

\bibitem{Altmannshofer:2019xda}
W.~Altmannshofer, J.~Davighi and M.~Nardecchia, {\it {Gauging the accidental
  symmetries of the standard model, and implications for the flavor
  anomalies}},  {\em Phys. Rev. D} {\bf 101} (2020), no.~1 015004
  [\href{http://arXiv.org/abs/1909.02021}{{\tt 1909.02021}}].

\bibitem{Capdevila:2020rrl}
B.~Capdevila, A.~Crivellin, C.~A. Manzari and M.~Montull, {\it {Explaining
  $b\to s\ell^+\ell^-$ and the Cabibbo angle anomaly with a vector triplet}},
  {\em Phys. Rev. D} {\bf 103} (2021), no.~1 015032
  [\href{http://arXiv.org/abs/2005.13542}{{\tt 2005.13542}}].

\bibitem{Davighi:2020qqa}
J.~Davighi, M.~Kirk and M.~Nardecchia, {\it {Anomalies and accidental
  symmetries: charging the scalar leptoquark under L$_{\mu}$ \ensuremath{-}
  L$_{\tau}$}},  {\em JHEP} {\bf 12} (2020) 111
  [\href{http://arXiv.org/abs/2007.15016}{{\tt 2007.15016}}].

\bibitem{Allanach:2020kss}
B.~C. Allanach, {\it {$U(1)_{B_3-L_2}$ explanation of the neutral current
  $B$\ensuremath{-}anomalies}},  {\em Eur. Phys. J. C} {\bf 81} (2021), no.~1
  56 [\href{http://arXiv.org/abs/2009.02197}{{\tt 2009.02197}}]. [Erratum:
  Eur.Phys.J.C 81, 321 (2021)].

\bibitem{Bednyakov:2021fof}
A.~Bednyakov and A.~Mukhaeva, {\it {Flavour Anomalies in a $U(1)$ SUSY
  Extension of the SM}},  {\em Symmetry} {\bf 13} (2021), no.~2 191.

\bibitem{Davighi:2021oel}
J.~Davighi, {\it {Anomalous $Z^\prime$ bosons for anomalous $B$ decays}},
  \href{http://arXiv.org/abs/2105.06918}{{\tt 2105.06918}}.

\bibitem{Greljo:2021npi}
A.~Greljo, Y.~Soreq, P.~Stangl, A.~E. Thomsen and J.~Zupan, {\it {Muonic Force
  Behind Flavor Anomalies}},  \href{http://arXiv.org/abs/2107.07518}{{\tt
  2107.07518}}.

\bibitem{Wang:2021uqz}
X.~Wang, {\it {Muon $(g-2)$ and Flavor Puzzles in the $U(1)^{}_{X}$-gauged
  Leptoquark Model}},  \href{http://arXiv.org/abs/2108.01279}{{\tt
  2108.01279}}.

\bibitem{ParticleDataGroup:2020ssz}
{\bf Particle Data Group} Collaboration, P.~A. Zyla {\em et.~al.}, {\it {Review
  of Particle Physics}},  {\em PTEP} {\bf 2020} (2020), no.~8 083C01.

\bibitem{Allanach:2021kzj}
B.~C. Allanach, J.~E. Camargo-Molina and J.~Davighi, {\it {Global fits of third
  family hypercharge models to neutral current B-anomalies and electroweak
  precision observables}},  {\em Eur. Phys. J. C} {\bf 81} (2021), no.~8 721
  [\href{http://arXiv.org/abs/2103.12056}{{\tt 2103.12056}}].

\bibitem{Allanach:2019mfl}
B.~C. Allanach, J.~M. Butterworth and T.~Corbett, {\it {Collider constraints on
  Z$^{\prime}$ models for neutral current B-anomalies}},  {\em JHEP} {\bf 08}
  (2019) 106 [\href{http://arXiv.org/abs/1904.10954}{{\tt 1904.10954}}].

\bibitem{ATLAS:2019erb}
{\bf ATLAS} Collaboration, G.~Aad {\em et.~al.}, {\it {Search for high-mass
  dilepton resonances using 139 fb$^{-1}$ of $pp$ collision data collected at
  $\sqrt{s}=$13 TeV with the ATLAS detector}},  {\em Phys. Lett. B} {\bf 796}
  (2019) 68--87 [\href{http://arXiv.org/abs/1903.06248}{{\tt 1903.06248}}].

\bibitem{Christensen_2009}
N.~D. Christensen and C.~Duhr, {\it Feynrules ? feynman rules made easy},  {\em
  Computer Physics Communications} {\bf 180} (Sep, 2009) 1614?1641.

\bibitem{Degrande_2012}
C.~Degrande, C.~Duhr, B.~Fuks, D.~Grellscheid, O.~Mattelaer and T.~Reiter, {\it
  Ufo ? the universal feynrules output},  {\em Computer Physics Communications}
  {\bf 183} (Jun, 2012) 1201?1214.

\bibitem{Alwall_2014}
J.~Alwall, R.~Frederix, S.~Frixione, V.~Hirschi, F.~Maltoni, O.~Mattelaer,
  H.-S. Shao, T.~Stelzer, P.~Torrielli and M.~Zaro, {\it The automated
  computation of tree-level and next-to-leading order differential cross
  sections, and their matching to parton shower simulations},  {\em Journal of
  High Energy Physics} {\bf 2014} (Jul, 2014).

\bibitem{Sjostrand:2014zea}
T.~Sj\"ostrand, S.~Ask, J.~R. Christiansen, R.~Corke, N.~Desai, P.~Ilten,
  S.~Mrenna, S.~Prestel, C.~O. Rasmussen and P.~Z. Skands, {\it {An
  introduction to PYTHIA 8.2}},  {\em Comput. Phys. Commun.} {\bf 191} (2015)
  159--177 [\href{http://arXiv.org/abs/1410.3012}{{\tt 1410.3012}}].

\bibitem{2007}
J.~Alwall, S.~H\"oche, F.~Krauss, N.~Lavesson, L.~L\"onnblad, F.~Maltoni,
  M.~Mangano, M.~Moretti, C.~Papadopoulos, F.~Piccinini and et~al., {\it
  Comparative study of various algorithms for the merging of parton showers and
  matrix elements in hadronic collisions},  {\em The European Physical Journal
  C} {\bf 53} (Dec, 2007) 473--500.

\bibitem{deFavereau:2013fsa}
{\bf DELPHES 3} Collaboration, J.~de~Favereau, C.~Delaere, P.~Demin,
  A.~Giammanco, V.~Lema\^\i{}tre, A.~Mertens and M.~Selvaggi, {\it {DELPHES 3,
  A modular framework for fast simulation of a generic collider experiment}},
  {\em JHEP} {\bf 02} (2014) 057 [\href{http://arXiv.org/abs/1307.6346}{{\tt
  1307.6346}}].

\bibitem{Cacciari:2011ma}
M.~Cacciari, G.~P. Salam and G.~Soyez, {\it {FastJet User Manual}},  {\em Eur.
  Phys. J. C} {\bf 72} (2012) 1896 [\href{http://arXiv.org/abs/1111.6097}{{\tt
  1111.6097}}].

\bibitem{Cacciari:2005hq}
M.~Cacciari and G.~P. Salam, {\it {Dispelling the $N^{3}$ myth for the $k_t$
  jet-finder}},  {\em Phys. Lett. B} {\bf 641} (2006) 57--61
  [\href{http://arXiv.org/abs/hep-ph/0512210}{{\tt hep-ph/0512210}}].

\bibitem{Cacciari:2008gp}
M.~Cacciari, G.~P. Salam and G.~Soyez, {\it {The anti-$k_t$ jet clustering
  algorithm}},  {\em JHEP} {\bf 04} (2008) 063
  [\href{http://arXiv.org/abs/0802.1189}{{\tt 0802.1189}}].

\bibitem{CMS:2021ctt}
{\bf CMS} Collaboration, A.~M. Sirunyan {\em et.~al.}, {\it {Search for
  resonant and nonresonant new phenomena in high-mass dilepton final states at
  $ \sqrt{s} $ = 13 TeV}},  {\em JHEP} {\bf 07} (2021) 208
  [\href{http://arXiv.org/abs/2103.02708}{{\tt 2103.02708}}].

\bibitem{ATLAS:2021mla}
{\bf ATLAS} Collaboration, G.~Aad {\em et.~al.}, {\it {Search for new phenomena
  in final states with two leptons and one or no $b$-tagged jets at $\sqrt{s} =
  13$ TeV using the ATLAS detector}},
  \href{http://arXiv.org/abs/2105.13847}{{\tt 2105.13847}}.

\bibitem{ATLAS:2019fgd}
{\bf ATLAS} Collaboration, G.~Aad {\em et.~al.}, {\it {Search for new
  resonances in mass distributions of jet pairs using 139 fb$^{-1}$ of $pp$
  collisions at $\sqrt{s}=13$ TeV with the ATLAS detector}},  {\em JHEP} {\bf
  03} (2020) 145 [\href{http://arXiv.org/abs/1910.08447}{{\tt 1910.08447}}].

\bibitem{CMS:2019gwf}
{\bf CMS} Collaboration, A.~M. Sirunyan {\em et.~al.}, {\it {Search for high
  mass dijet resonances with a new background prediction method in
  proton-proton collisions at $\sqrt{s} =$ 13 TeV}},  {\em JHEP} {\bf 05}
  (2020) 033 [\href{http://arXiv.org/abs/1911.03947}{{\tt 1911.03947}}].

\bibitem{Aaboud_2018}
M.~Aaboud, G.~Aad, B.~Abbott, O.~Abdinov, B.~Abeloos, S.~H. Abidi, O.~S.
  AbouZeid, N.~L. Abraham, H.~Abramowicz and et~al., {\it Search for additional
  heavy neutral higgs and gauge bosons in the ditau final state produced in 36
  fb$^{-1}$ of pp collisions at $\sqrt{s}=13$ tev with the atlas detector},
  {\em Journal of High Energy Physics} {\bf 2018} (Jan, 2018).

\bibitem{ATLAS:2020lks}
{\bf ATLAS} Collaboration, G.~Aad {\em et.~al.}, {\it {Search for $
  t\overline{t} $ resonances in fully hadronic final states in $pp$ collisions
  at $ \sqrt{s} $ = 13 TeV with the ATLAS detector}},  {\em JHEP} {\bf 10}
  (2020) 061 [\href{http://arXiv.org/abs/2005.05138}{{\tt 2005.05138}}].

\bibitem{Allanach:2021gmj}
B.~C. Allanach, J.~M. Butterworth and T.~Corbett, {\it {Large Hadron Collider
  Constraints on Some Simple $Z'$ Models for $b\to s\mu^+\mu^-$ Anomalies}},
  \href{http://arXiv.org/abs/2110.13518}{{\tt 2110.13518}}.

\bibitem{Allanach:2019iiy}
B.~C. Allanach and J.~Davighi, {\it {Naturalising the third family hypercharge
  model for neutral current $B$-anomalies}},  {\em Eur. Phys. J. C} {\bf 79}
  (2019), no.~11 908 [\href{http://arXiv.org/abs/1905.10327}{{\tt
  1905.10327}}].

\end{thebibliography}\endgroup

\end{document}